\documentclass[final,onefignum,onetabnum]{siamart171218}
\usepackage{amsmath}
\DeclareMathOperator*{\argmin}{argmin} 
\usepackage{latexsym}
\usepackage{epsfig}
\usepackage{verbatim}
\usepackage{mathrsfs}
\usepackage{float}
\usepackage{bigints}
\usepackage{graphics}
\usepackage{lscape}
\usepackage{enumitem}
\usepackage{xcolor}

\usepackage{lipsum}
\usepackage{amsfonts}
\usepackage{graphicx}
\usepackage{epstopdf}
\usepackage{algorithmic}
\ifpdf
  \DeclareGraphicsExtensions{.eps,.pdf,.png,.jpg}
\else
  \DeclareGraphicsExtensions{.eps}
\fi

\newsiamremark{remark}{Remark}
\newsiamremark{example}{Example}
\newsiamremark{hypothesis}{Hypothesis}
\crefname{hypothesis}{Hypothesis}{Hypotheses}
\newsiamthm{claim}{Claim}

\headers{Fisher information for molecules}{M. R. Vahid, B. Hanzon, R. J. Ober}

\title{Fisher information matrix for single molecules with stochastic trajectories \thanks{Submitted to the editors February 2, 2019.
\funding{This work was supported in part by the National Institutes of Health (R01 GM085575).}}}

\author{Milad R. Vahid \footnotemark[5] {\thanks{D\lowercase{epartment of} B\lowercase{iomedical} E\lowercase{ngineering}, T\lowercase{exas} A\&M U\lowercase{niversity}, C\lowercase{ollege} S\lowercase{tation}, TX 77843, USA 
  \lowercase{(\email{milad.rafiee@tamu.edu}, \email{raimund.ober@tamu.edu}).}}} \thanks{Department of Biomedical Data Science, Stanford University, Stanford, CA 94305, USA (\email{miladrv@stanford.edu}).}
\and Bernard Hanzon\thanks{Edgeworth Centre for Financial Mathematics, School of Mathematical Sciences, University College Cork, Ireland
  (\email{b.hanzon@ucc.ie}).}
\and Raimund J. Ober \footnotemark[2] {\thanks{D\lowercase{epartment of} M\lowercase{olecular and} C\lowercase{ellular} M\lowercase{edicine}, T\lowercase{exas} A\&M H\lowercase{ealth} S\lowercase{cience} C\lowercase{enter}, C\lowercase{ollege} S\lowercase{tation}, TX 77843, USA \lowercase{(\email{milad.rafiee@tamu.edu}, \email{raimund.ober@tamu.edu})}.}} \thanks{Center for Cancer Immunology, Faculty of Medicine, University of Southampton, Southampton, UK (\email{r.ober@soton.ac.uk}).}}

\usepackage{amsopn}

\ifpdf
\hypersetup{
  pdftitle={Fisher information for molecules},
  pdfauthor={M. R. Vahid, B. Hanzon, R. J. Ober}
}
\fi

\begin{document}

\maketitle

\begin{abstract}
Tracking of objects in cellular environments has become a vital tool in molecular cell biology. A particularly important example is single molecule tracking which enables the study of the motion of a molecule in cellular environments by locating the molecule over time and provides quantitative information on the behavior of individual molecules in cellular environments, which were not available before through bulk studies. Here, we consider a dynamical system where the motion of an object is modeled by stochastic differential equations (SDEs), and measurements are the detected photons emitted by the moving fluorescently labeled object, which occur at discrete time points, corresponding to the arrival times of a Poisson process, in contrast to equidistant time points which have been commonly used in the modeling of dynamical systems. The measurements are distributed according to the optical diffraction theory, and therefore, they would be modeled by different distributions, e.g., an Airy profile for an in-focus and a Born and Wolf profile for an out-of-focus molecule with respect to the detector. For some special circumstances, Gaussian image models have been proposed. In this paper, we introduce a stochastic framework in which we calculate the maximum likelihood estimates of the biophysical parameters of the molecular interactions, e.g., diffusion and drift coefficients. More importantly, we develop a general framework to calculate the Cram\'er-Rao lower bound (CRLB), given by the inverse of the Fisher information matrix, for the estimation of unknown parameters and use it as a benchmark in the evaluation of the standard deviation of the estimates. There exists no established method, even for Gaussian measurements, to systematically calculate the CRLB for the general motion model that we consider in this paper. We apply the developed methodology to simulated data of a molecule with linear trajectories and show that the standard deviation of the estimates matches well with the square root of the CRLB. We also show that equally sampled and Poisson distributed time points lead to significantly different Fisher information matrices. 
\end{abstract}



\section{Introduction}
The ability to track objects of interest, e.g., subcellular organelles and molecules, in cellular environments plays an important role in studying biological systems. In particular, single molecule tracking, which enables following subcellular processes at the single molecule level, has become a vital tool in cell biology \cite{saxton1,saxton2,saxton_new}. Traditionally, microscopy studies were bulk studies and the information from such studies reflected the behavior of ensembles of molecules as opposed to individual ones \cite{thesis}. Single molecule microscopy techniques have revolutionized the field of microscopy by providing quantitative information on the behavior of individual molecules in cellular environments, which were not available before through bulk studies \cite{2004,2006}. In biological studies, single molecule tracking methods have been used to study the intracellular trafficking of fluorescently labeled antibodies, e.g., prostate-specific membrane antigen (PSMA) antibodies \cite{psma1,psma2,psma3}, by analyzing the velocity and path of the fluorescent molecules. 

In general, the motion of an object in cellular environments is subject to different types of forces, e.g., deterministic forces due to the environment and random forces due to random collisions with other objects \cite{Schuss,kervrann}. It has been shown that the motion of a moving object in such environments can be modeled by stochastic differential equations (SDEs) \cite{fokker}. In particular, in many biological applications, solutions of linear SDEs are good fits to experimental single molecule trajectories \cite{calderon1,calderon2,calderon3}. In a basic fluorescence microscope, a fluorescently labeled object of interest is imaged by a detector which detects the photons emitted by the object during the acquisition time. Since the detection process of the emitted photons is inherently a random phenomenon, the acquired measurements are stochastic in nature. These measurements, according to the optical diffraction theory, can be modeled by different distributions. For example, a typical distribution for an in-focus molecule is an Airy profile \cite{jerry}, whereas, classical Born and Wolf profiles \cite{born} are used instead for out-of-focus molecules. In some cases, it is possible and computationally beneficial to approximate these complex profiles with simple Gaussian models \cite{anish}. 

In many dynamical systems, the time points of the measurements are assumed to be equidistant. However, the time points of detection of the photons correspond to the arrival times of a Poisson process \cite{2004,2006}. This gives rise to non-uniform sampling of the continuous-time stochastic process that describes the motion of the object. Since the parameters of the motion model of the object are highly time-dependent, this randomized non-uniform sampling causes significant fluctuations in the values of these parameters.

In recent years, many methods have been developed to analyze the trajectories of a molecule in cellular environments. In most of these methods, the model for the motion of the molecule is assumed to be limited to a Brownian motion (pure diffusion) model described only by the diffusion coefficient, and only few of the available methods consider more general motion models. The methods developed to analyze pure diffusion models are mostly based on the mean square displacement approach \cite{Qian}, in which the diffusion coefficient is estimated by a linear regression of the mean square displacement of the Gaussian distributed observed locations of the molecule as a function of the time lag \cite{berglund2010,berglund2012,Michalet2010}. Mean square displacement-based methods are not the only approaches used to estimate the diffusion coefficient from a set of measurements. For example, Relich et al. \cite{Lidke} have proposed a method for the maximum likelihood estimation of the diffusion coefficient, with an information-based confidence interval, from Gaussian measurements. In all of these methods, the motion of a molecule is assumed as a pure diffusion model, and the measurements are modeled by independent and identically distributed Gaussian random variables \cite{review2015}. 

However, in general, the motion of a molecule is not limited to the pure diffusion model, and the diffusion coefficient is only one of the parameters that play a role in the motion of the molecule. Also, the Gaussian assumption for the measurements is problematic in practice due to the fact that the Gaussian model is often not an accurate analytical model. In \cite{andersson}, Ashley and Andersson have proposed a simultaneous localization and parameter estimation algorithm for more complex motion models, such as confined \cite{fokker} and tethered motions \cite{tetered}, which employs the expectation maximization algorithm in conjunction with sequential Monte Carlo methods \cite{automatica}. For the general object tracking problem, in \cite{singh1,singh2}, a sequential Monte Carlo method has been developed for the parameter estimation from nonlinear non-Gaussian state-space models. Briane et al. \cite{kervrann} have developed a method for classifying the object trajectories in living cells into three types of diffusion: Brownian motion, subdiffusion (diffusion in a closed domain or in a crowded area) and superdiffusion (diffusion in a specific direction). In \cite{calderon1,calderon2,calderon3}, the motion of a moving object has been described more generally by a linear SDE, and the parameters of the model has been estimated using a maximum likelihood estimation method. However, they do not consider randomness of the time points at which the measurements occur. Their proposed framework also does not allow for non-Gaussian measurements.

In this paper, we address the above limitations by considering a more general dynamical system with arbitrary distributed measurements, which occur at Poisson distributed time points, that allows for more general motion models for an object of interest. Here, the motion of an object in cellular environments is modeled by stochastic differential equations, and the measurements are the detected photons emitted by the moving fluorescently labeled object. As mentioned earlier, these measurements can be modeled by non-Gaussian distributions. We develop a stochastic framework in which we calculate the maximum likelihood estimates of the biophysical parameters of the molecular interactions, e.g., diffusion and drift coefficients. 

According to a well-known result from estimation theory, assuming that the estimator is unbiased, its standard deviation is then at best equal to the square root of the CRLB, which is given by the inverse of the Fisher information matrix \cite{2004,2006,jerry}. More importantly, in order to evaluate the performance of our proposed estimation method, we develop a general framework to calculate the Fisher information matrix of the unknown parameters of the general motion model. There are some cases in which Gaussian approximations of measurements are very useful due to, for example, the ability of using computationally efficient algorithms in linear systems or the Kalman filter formulae. In particular, for Gaussian measurements, we calculate the Fisher information matrix by taking advantage of its relationship with the Kalman filter formula through a computationally efficient algorithm. To the best of our knowledge, even for Gaussian measurements, there currently exists no systematic methodology to evaluate the standard deviations of the estimates using the CRLB for the general motion model considered here. 

To assess the performance of the proposed estimation method, we apply it to simulated data sets comprising linear trajectories of a molecule with Gaussian, Airy and classical model of Born and Wolf measurements. The results show that there is no systematic bias associated with the method. In addition, we show that the means of the distributions of the prediction of the molecule locations are able to follow the true locations of the molecule for the all different types of measurements. In particular, for data sets comprising repeat trajectories of a molecule with Gaussian measurements, it is shown that the standard deviations of the diffusion and drift estimates are close to the square roots of their corresponding CRLBs. We also show that, in case that we have one detected photon, the Fisher information matrices obtained for an Airy and its corresponding approximating Gaussian profile are different from each other, and therefore, the use of the Gaussian approximation can be problematic in some applications. We show that equally sampled time points, which have been commonly used in most dynamical systems, and Poisson distributed time points can lead to significantly different Fisher information matrices. We further show that even the results obtained for different realizations of a Poisson process can vary notably. 

This paper is organized as follows. In Section \ref{fundamentalsection}, we present the statistical description of the acquired data, and derive a general formula for the likelihood function of the described data model. Section \ref{linear sde} is devoted to introduce linear stochastic systems and calculate the likelihood function in case that the object is undergoing this type of trajectories. In Section \ref{mle}, we propose a mathematical framework to calculate the maximum likelihood estimates of the parameters of interest, such as the parameters of the motion model of the molecule. Section \ref{crlbsection} is devoted to calculate general expressions for the CRLB and Fisher information matrix relating to the parameter estimation problem.

In this paper, we use the following notation
\begin{align}
\label{set}
\mathcal{C}^l\times\mathbb{R}_{[t]}^l:=\left\{\left(r_1,\cdots,r_l,\tau_1,\cdots,\tau_l\right)|r_1,\cdots,r_l\in\mathcal{C},t_0\leq\tau_1<\cdots<\tau_l\leq t\right\},
\end{align}
where $\mathcal{C}:=\mathbb{R}^2$, $t_0\in\mathbb{R}$, and $l=1,2,\cdots$. If there is no bound on $\tau_l$, we denote the set in Eq. (\ref{set}) by $\mathcal{C}^l\times\mathbb{R}_{[\infty]}^l$.

\section{Fundamental data model}
\label{fundamentalsection}

\begin{figure}[htbp]
\centering\includegraphics[width=0.85\textwidth]{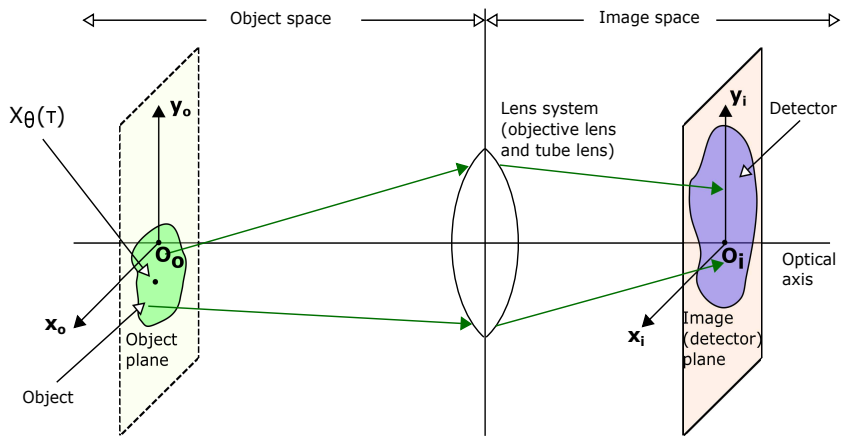}
\DeclareGraphicsExtensions{.eps} \caption{ Schematic of an optical microscope.
An object located in the object (focal) plane is imaged by an optical lens system and the image of the object is acquired by the planar detector in the image space. A 3D random variable $X_{\theta}(\tau), \tau\geq t_0$, describes the location of the object in the object plane at time $\tau$.}
\label{drawing}
\end{figure}

A basic setup of an optical system considered here is shown in Fig. \ref{drawing}, where an object is in the object space and its image is captured by a planar detector in the image space. In the fundamental data model, we assume that the microscopy image data is acquired under ideal conditions. It assumes the use of an image detector that has an unpixelated photon detection area. The detection of a photon is intrinsically random in terms of both the time and the location on the detector at which the photon is detected. In general, the temporal part of the detection of the emitted photons can be modeled as a counting process $\left\{N(\tau), \tau\geq t_0\right\}$. Here, we assume that $\left\{N(\tau), \tau\geq t_0\right\}$ is a Poisson process referred to as the \emph{photon detection process} that is characterized by the intensity function $\Lambda(\tau), \tau\geq t_0$, referred to as \emph{the photon detection rate}. The spatial component of the photon detection process is specified by random variables, referred to as \emph{the photon location variables}, that describe the locations at which photons emitted by the object of interest are detected.
 
In the following definition, we define a spatio-temporal process referred to as the \emph{image detection process}, which models the acquired data, for two different acquisition methods, one when the time interval over which photons are detected is given and the other when the total number of detected photons is given. For a fixed acquisition time, due to the stochastic nature of photon emission, the total number of detected photons varies for every image, while in the other case, the number of detected photons remains the same. 

\begin{definition}
\label{def1}

Let $\mathcal{C}:=\mathbb{R}^2$ denote a non-pixelated detector. Let $\mathbb{R}^n, n=1,2,\cdots$, be the $n$-dimensional full parameter space. Let the parameter space $\Theta$ describe an open subset of $\mathbb{R}^n$ containing the true parameters. Elements in $\Theta$ are described by a parameter vector $\theta\in\Theta$. Let the one-dimensional (1D) random variables $T_1,T_2,\cdots$, describe the time points of detection of the photons that impact the detector $\mathcal{C}$, which are arrival time points associated with a Poisson process with intensity function $\Lambda(\tau), \tau\geq t_0, t_0\in\mathbb{R}$. Let $U_1,U_2,\cdots$, be 2D random variables that describe the locations of detection of the photons that impact the detector $\mathcal{C}$. For $l=1,2,\cdots$, let $\mathcal{U}_l:=\left(U_1,\cdots,U_l\right), \mathcal{U}_0=\emptyset$, and $\mathcal{T}_l:=\left(T_1,\cdots,T_l\right), \mathcal{T}_0=\emptyset$. Assume that the current location of the detected photon, given the current and previous time points, is independent of the future time points, i.e., for $r\in\mathcal{C}$ and $t_0\leq\tau_1<\tau_2<\cdots$,  
\begin{align*}
p_{U_l|\mathcal{T}_k}\Big(r|\tau_1,\cdots,\tau_k\Big)=p_{U_l|\mathcal{T}_l}\Big(r|\tau_1,\cdots,\tau_l\Big),\quad \mbox{for all}\ k,l=1,2,\cdots,\quad k\geq l,
\end{align*}
where, for random vectors $X$ and $Y$, the conditional probability density function of $X$, given $Y$, is denoted by $p_{X|Y}$. In other words, we assume that it may depend on past and current inputs but not future inputs. This assumption is natural in the context of the modeling of the dynamics of biomolecular processes such as the stochastic trajectory of a single molecule or organelle in a cellular context, where future effects do not impact the present.

1. For a fixed acquisition time interval $[t_0,t]$, an \emph{image detection process\\
$\mathcal{G}_{[t]}\Big(\left(\mathcal{U}_{[t]},\mathcal{T}_{[t]}\right),\mathcal{C},\Theta\Big)$ for a time interval $[t_0,t]$} is defined as a spatio-temporal process whose temporal part $\mathcal{T}_{[t]}$ and spatial part $\mathcal{U}_{[t]}$ describe the time points and the locations of detection of the photons that impact the detector $\mathcal{C}$ in the time interval $[t_0,t]$, respectively, i.e., for $\omega\in\Omega$, where $\Omega$ is the sample space, 
\begin{align*}
U_{S_t(\omega)}=T_{S_t(\omega)}=\emptyset,\quad S_t(\omega)=0,
\end{align*}
and
\begin{small}
\begin{align*}
\mathcal{T}_{[t]}(\omega):=\left(T_1(\omega),\cdots,T_{S_t(\omega)}(\omega)\right),\quad \mathcal{U}_{[t]}(\omega):=\left(U_1(\omega),\cdots,U_{S_t(\omega)}(\omega)\right),\quad S_t(w)> 0,
\end{align*}
\end{small}
where $t_0\leq T_1(\omega)<\cdots<T_{S_t(\omega)}(\omega)\leq t$, and $S_t$ is a discrete 1D random variable that takes its values in the non-negative integers such that $T_{S_t(\omega)}(\omega)\leq t, T_{S_t(\omega)+1}(\omega)> t, S_t(w)> 0$.

2. Given a fixed number $L=1,2,\cdots$, of photons, an \emph{image detection process $\mathcal{G}_L\Big(\left(\mathcal{U}_L,\mathcal{T}_L\right),\mathcal{C},\Theta\Big)$ for a fixed number $L$ of photons} is defined as a spatio-temporal process whose temporal and spatial parts describe the time points and the locations of detection of the $L$ photons that impact the detector $\mathcal{C}$, respectively. Moreover, given $\mathcal{T}_L=\left(\tau_1,\cdots,\tau_L\right), t_0\leq\tau_1<\tau_2<\cdots<\tau_L$, $\mathcal{G}_{\tau_1,\cdots,\tau_L}\Big(\left(\mathcal{U}_L,\mathcal{T}_L\right),\mathcal{C},\Theta\Big)$ is referred to as the \emph{image detection process at fixed time points $\tau_1,\cdots,\tau_L$}. 
\end{definition}

In Theorem \ref{main1}, we state expressions for the probability/probability density functions of image detection processes for a fixed time interval and for a fixed number of photons in terms of the conditional distributions of the locations of the detected photons, given the previous locations and the current and previous time points of the detected photons. We further show that each of these conditional distributions can be expressed in terms of a scaled and shifted version of the image of the object and the distribution of the prediction of the object location, given the previous locations and time points of the detected photons. All the proofs in the paper are placed in the supplementary material. We drop the parameter vector $\theta\in\Theta$, when it is clear from the context.

\begin{theorem}
\label{main1}
Let $\mathcal{G}_{[t]}\Big(\left(\mathcal{U}_{[t]},\mathcal{T}_{[t]}\right),\mathcal{C},\Theta\Big)$ and $\mathcal{G}_L\Big(\left(\mathcal{U}_L,\mathcal{T}_L\right),\mathcal{C},\Theta\Big)$ be image detection processes for a time interval $[t_0,t]$ and for a fixed number $L$ of photons, respectively. Let $\mathcal{D}_{[t]}:=\left(\mathcal{U}_{[t]},\mathcal{T}_{[t]}\right), \mathcal{D}_k:=\left(\mathcal{U}_k,\mathcal{T}_k\right), k=0,1,\cdots$.
 
1. Then, the probability of $\mathcal{D}_{[t]}=\emptyset$ and $N(t)=0$ is given by
\begin{align*}
P\Big(\mathcal{D}_{[t]}=\emptyset,N(t)=0\Big)=e^{-\int_{t_0}^t\Lambda(\tau)d\tau},
\end{align*}
and the probability density function $p_{[t]}$ of $\mathcal{D}_{[t]}$ and $N(t)$ is given by 
\begin{small}
\begin{align}
\label{likelihood000}
p_{[t]}\Big(d_K,K\Big)=e^{-\int_{t_0}^t\Lambda(\tau)d\tau}\prod_{k=1}^K\Lambda(\tau_k)\left[\prod_{l=1}^Kp_{U_l|T_l,\mathcal{D}_{l-1}}\Big(r_l|\tau_l,d_{l-1}\Big)\right],
\end{align}
\end{small}
where $d_K\in\mathcal{C}^K\times\mathbb{R}_{[t]}^K, K=1,2,\cdots$, and $p_{U_l|T_l,\mathcal{D}_{l-1}}$ denotes the conditional probability density function of $U_l$, given $T_l, \mathcal{D}_{l-1}$, with $p_{U_1|T_1,\mathcal{D}_0}\Big(r_1|\tau_1,d_0\Big):=p_{U_1|T_1}\Big(r_1|\tau_1\Big)$.

2. Moreover, the probability density function $p_L$ of $\mathcal{D}_L$ is given by
\begin{small}
\begin{align}
\label{likelihood001}
p_L\Big(d_L\Big)=e^{-\int_{t_0}^{\tau_L}\Lambda(\tau)d\tau}\prod_{k=1}^L\Lambda(\tau_k)\left[\prod_{l=1}^Lp_{U_l|T_l,\mathcal{D}_{l-1}}\Big(r_l|\tau_l,d_{l-1}\Big)\right],\quad d_L\in\mathcal{C}^L\times\mathbb{R}_{[\infty]}^L.
\end{align}
\end{small}
\end{theorem}

\begin{proof}
See Section \ref{proof:main1} in the supplementary material.
\end{proof}

Note that, as can be seen in the above theorem, the probability density function of an image detection process for a time interval $[t_0,t]$ depends on the integral of the photon detection rate $\Lambda(\tau), \tau\geq t_0$, over the time interval $[t_0,t]$, and the probability density function of an image detection process for a fixed number $L$ of photons depends on the integral of the photon detection rate over the time interval $[t_0,\tau_L]$, where $\tau_L$ denotes the time point of the $L^{th}$ (last) detected photon. 

The probability density function of the location at which a photon emitted by the object of interest is detected, is referred to as the \emph{image profile} of the object. So far we have made no assumptions about the specific functional form of the image profile of the object. In many practical cases, the image profile can be described as a scaled and shifted version of the image function. In such cases, an image function describes the image of an object on the detector plane at unit lateral magnification. Also, in general, the trajectory of the object can be described by a random process. In the following definition, we define image detection processes driven by a stochastic trajectory of the object and the image function for a fixed time interval and for a fixed number of photons.

\begin{definition}
\label{def2}

Let $\mathcal{G}_{[t]}\Big(\left(\mathcal{U}_{[t]},\mathcal{T}_{[t]}\right),\mathcal{C},\Theta\Big)$ and $\mathcal{G}_L\Big(\left(\mathcal{U}_L,\mathcal{T}_L\right),\mathcal{C},\Theta\Big)$ be image detection processes for a time interval $[t_0,t]$ and for a fixed number $L$ of photons, respectively. Let $X(\tau), \tau\geq t_0$, denote a 3D random process that describes the 3D stochastic trajectory of the object. Also, let $\left\{f_x\right\}_{x\in\mathbb{R}^3}$ defined on the detector $\mathcal{C}$, be a family of image profiles of an object located at $x\in\mathbb{R}^3$ in the object space. Assume that the current location of the detected photon, given the current location of the object, is independent of the previous locations and time points of the detected photons, i.e., for all $x\in\mathbb{R}^3$,
\begin{align*}
p_{U_l|X(T_l),T_l,\mathcal{D}_{l-1}}\Big(r_l|x,\tau_l,d_{l-1}\Big)=p_{U_l|X(\tau_l)}\Big(r_l|x\Big):=f_x\left(r_l\right),\quad r_l\in\mathcal{C},
\end{align*}
where $d_l\in\mathcal{C}^l\times\mathbb{R}_{[t]}^l$ for $\mathcal{G}_{[t]}$, $d_l\in\mathcal{C}^l\times\mathbb{R}_{[\infty]}^l$ for $\mathcal{G}_L$, $p_{U_l|X(T_l),T_l,\mathcal{D}_{l-1}}$ is the conditional probability density function of $U_l$, given $X(T_l),T_l,\mathcal{D}_{l-1}$, and $p_{U_l|X(\tau_l)}$ denotes the conditional probability density function of $U_l$, given $X(\tau_l)$. This assumption is justified as the process of the image formation, photon emission etc. only depends on the position of the emitting fluorescent object at the particular point in time and not on prior events.

Assume that there exists a function $q_{z_0}{:}\ \mathbb{R}^2\mapsto\mathbb{R}, z_0\in\mathbb{R}$, such that for an invertible matrix $M\in\mathbb{R}^{2\times 2}$ and $x:=\left(x_0,y_0,z_0\right)\in\mathbb{R}^3$,
\begin{align}
\label{imagefunctionmain}
f_x\left(r\right):=\frac{1}{\left|\det\left(M\right)\right|}q_{z_0}\Big(M^{-1}r-(x_0,y_0)^T\Big),\quad r\in\mathcal{C}.
\end{align}
In the above equation, $q_{z_0}$, which is referred to as the \emph{image function}, is a function that describes, at unit lateral magnification, the image of the object in the detector plane when the object is located at $\left(0,0,z_0\right)$ in the object space.

\emph{Image detection processes $\mathcal{G}_{[t]}\Big(X,\left(\mathcal{U}_{[t]},\mathcal{T}_{[t]}\right),q,\mathcal{C},\Theta\Big)$ and $\mathcal{G}_L\Big(X,\left(\mathcal{U}_L,\mathcal{T}_L\right),q,\mathcal{C},\\
\Theta\Big)$ driven by the stochastic trajectory $X$ and image function $q$ for a time interval $[t_0,t]$ and for a fixed number $L$ of photons} are defined as the spatio-temporal processes $\mathcal{G}_{[t]}$ and $\mathcal{G}_L$, respectively.
\end{definition}

In the classical case of a measurement error, the image function $q_{z_0}$ is defined as a function of $\left(r-M(x_0,y_0)^T\right)$, which is the deviation between two locations in the image space. Here, however, in order to be consistent with our previous framework developed for a static object, $q_{z_0}$ is defined as a function of $\left(M^{-1}r-(x_0,y_0)^T\right)$, which is the difference between two points in the object space.

\begin{figure}[htbp]
\centering\includegraphics[width=1\textwidth]{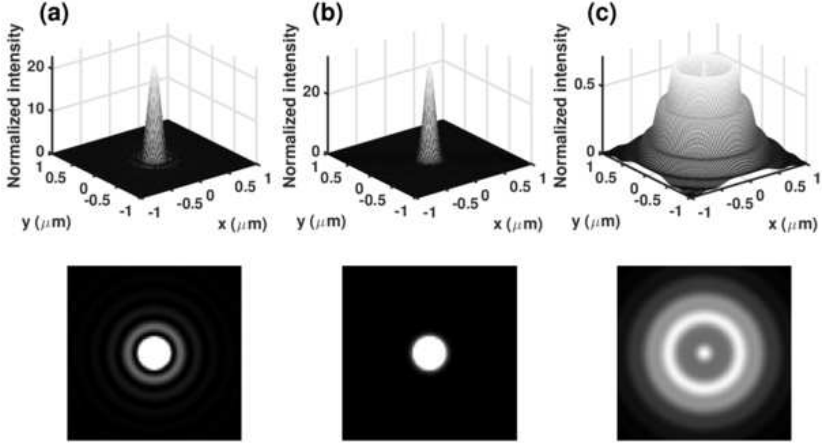}
\DeclareGraphicsExtensions{.eps} \caption{Image function examples. (a) Airy and (b) symmetric Gaussian profiles, which describe the images of an in-focus point source, simulated by Eqs. (\ref{airyprofileeq}) and (\ref{gaussianimagefunction}), respectively, with the parameters given in Section \ref{parameters}. (c) Born and Wolf profile simulated by Eq. (\ref{bornimagefunction}) with the out-of-focus level $z_0=1$ $\mu$m, and the parameters given in Section \ref{parameters}.}
\label{imagefunction}
\end{figure}

We next illustrate specific image functions that describe the image of a point source. According to the optical diffraction theory, when a point source is in-focus with respect to the detector, the intensity distribution of the image of the point source is described by an Airy profile given by \cite{2006} (see Fig. \ref{imagefunction}(a))
\begin{align}
\label{airyprofileeq}
q(x,y)=\frac{J_1^2\left(\frac{2\pi n_a}{\lambda}\sqrt{x^2+y^2}\right)}{\pi\left(x^2+y^2\right)},\quad (x,y)\in\mathbb{R}^2,
\end{align}
where $n_a$ denotes the numerical aperture of the objective lens, $\lambda$ denotes the emission wavelength of the molecule, and $J_1$ denotes the first order Bessel function of the first kind. The 2D Gaussian profile, on the other hand, which has been widely used to approximate the Airy profile, is given by
\begin{align}
\label{gaussianimagefunction}
q(x,y)=\frac{1}{2\pi\sigma^2}e^{-\frac{1}{2}\left(\frac{x^2+y^2}{\sigma^2}\right)},\quad (x,y)\in\mathbb{R}^2,
\end{align}
where $\sigma>0$.

For an out-of-focus point source, the image function can be obtained by the classical Born and Wolf model given by \cite{born}
\begin{align}
\label{bornimagefunction}
q_{z_0}(x,y)=\frac{4\pi n_a^2}{\lambda^2}\left|\int_0^1J_0\left(\frac{2\pi n_a}{\lambda}\sqrt{x^2+y^2}\rho\right)e^{\frac{j\pi n_a^2z_0}{n_o\lambda}\rho^2}\rho d\rho\right|^2,\quad (x,y)\in\mathbb{R}^2,
\end{align}
where $J_0$ is the zeroth-order Bessel function of the first kind, $n_o$ is the refractive index of the objective lens immersion medium, and $z_0\in\mathbb{R}$ is the $z$-location of the point source on the optical axis in the object space. When the point source is in-focus with respect to the detector, i.e., it lies in the object plane, then $z_0=0$ and Eqs. (\ref{airyprofileeq}) and (\ref{bornimagefunction}) are equivalent.   

We calculate $p_{U_l|T_l,\mathcal{D}_{l-1}}, l=1,2,\cdots$, for more general cases. In the following corollary to Theorem \ref{main1}, by describing these conditional probability density functions in terms of the image function, we derive expressions for the probability density functions of the image detection processes driven by the stochastic trajectory $X$ and image function $q$ for a time interval $[t_0,t]$ and for a fixed number $L$ of photons.

\begin{corollary}
\label{corollary1}
Let $\mathcal{G}_{[t]}\Big(X,\left(\mathcal{U}_{[t]},\mathcal{T}_{[t]}\right),q,\mathcal{C},\Theta\Big)$ (or $\mathcal{G}_L\Big(X,\left(\mathcal{U}_L,\mathcal{T}_L\right),q,\mathcal{C},\Theta\Big)$) be an image detection process driven by the stochastic trajectory $X$ and image function $q$ for a time interval $[t_0,t]$ (or for a fixed number $L$ of photons). Then, the conditional probability density function $p_{U_l|T_l,\mathcal{D}_{l-1}}, l=1,2,\cdots$, in Eq. (\ref{likelihood000}) (or in Eq. (\ref{likelihood001})) of Theorem \ref{main1} is given by, for $x:=\left(x_0,y_0,z_0\right)\in\mathbb{R}^3$,
\begin{align}
\label{eqfixed2}
p_{U_l|T_l,\mathcal{D}_{l-1}}\Big(r_l|\tau_l,d_{l-1}\Big)&=\int_{\mathbb{R}^3}f_{x}\left(r_l\right)p_{pr_l}\Big(x|\tau_l,d_{l-1}\Big)dx\nonumber\\
&=\frac{1}{\left|\det(M)\right|}\int_{\mathbb{R}^3}q_{z_0}\Big(M^{-1}r_l-\left(x_0,y_0\right)\Big)p_{pr_l}\Big(x|\tau_l,d_{l-1}\Big)dx,
\end{align}
where $d_l\in\mathcal{C}^l\times\mathbb{R}_{[t]}^l$ (or $d_l\in\mathcal{C}^l\times\mathbb{R}_{[\infty]}^l$), $p_{pr_l}:=p_{X(T_l)|T_l,\mathcal{D}_{l-1}}$ denotes the distribution of the prediction of the object location, $p_{pr_1}\Big(x|\tau_1,d_0\Big):=p_{pr_1}\Big(x|\tau_1\Big)$, and $f_x, x\in\mathbb{R}^3$, is the image profile of an object located at $x$ in the object space.
\end{corollary}

\begin{proof}
See Section \ref{proof:corollary1} in the supplementary material.
\end{proof}

As can be seen in the above corollary, the expression of the probability density function of the image detection process depends on the distribution $p_{pr_l}, l=1,2,\cdots$, of the prediction of the object location, given the previous locations of the detected photons and the current and previous time points. In the following section, we introduce linear stochastic systems and calculate $p_{pr_l}, l=1,2,\cdots$, for them.

In Theorem \ref{main1}, we expressed the probability density functions of image detection processes in terms of conditional probability densities $p_{U_l|T_l,\mathcal{D}_{l-1}}, l=1,2,\cdots$, of the locations of the detected photons, given the previous locations and the current and previous time points of the detected photons. In particular, for an object with a deterministic trajectory or a static object, the conditional probability densities $p_{U_l|T_l,\mathcal{D}_{l-1}}, l=1,2,\cdots$, are given as follows. For an object with deterministic trajectory $X(\tau)\in\mathbb{R}^3,\tau\geq t_0$, we have
\begin{align}
\label{specific1}
p_{U_l|T_l,\mathcal{D}_{l-1}}\Big(r_l|\tau_l,d_{l-1}\Big)=p_{U_l|T_l}(r_l|\tau_l):=f_{X(\tau_l)}\left(r_l\right).
\end{align}
Also, for a static object with position $X_0\in\mathbb{R}^3$, we have
\begin{align}
\label{specific2}
p_{U_l|T_l,\mathcal{D}_{l-1}}\Big(r_l|\tau_l,d_{l-1}\Big)=p_{U_l}(r_l):=f_{X_0}\left(r_l\right).
\end{align}

\section{Linear stochastic systems}
\label{linear sde}

In general, the motion of an object in cellular environments is subject to different types of forces, e.g., deterministic forces due to the environment and random forces due to random collisions with other objects \cite{Schuss,kervrann}. The 3D random variable $X(\tau)$ denotes the location of the object at time $\tau\geq t_0$. Then, the motion of the object is assumed to be modeled through a general state space system with state $\tilde{X}(\tau)\in\mathbb{R}^k, \tau\geq t_0$, as 
\begin{align}
\label{sde1}
\tilde{X}(\tau_{l+1})=\tilde{\phi}(\tau_l,\tau_{l+1})\tilde{X}(\tau_l)+\tilde{W}(\tau_l,\tau_{l+1}),\quad \tau_0:=t_0\leq\tau_1<\cdots<\tau_{l+1}<\cdots,
\end{align}
where we assume that there exists a matrix $H\in\mathbb{R}^{3\times k}$ such that $X(\tau)=H\tilde{X}(\tau), \tau\geq t_0$, $\tilde{\phi}(\tau_l,\tau_{l+1})\in\mathbb{R}^{k\times k}$ is a state transition matrix, and $\left\{\tilde{W}(\tau_l,\tau_{l+1})\right.$, $\left. l=1,2,\cdots\right\}$ is a sequence of $k$-dimensional random variables with probability density functions $p_{\tilde{W}(\tau_l,\tau_{l+1})}$. We also assume that the initial state $\tilde{X}(t_0)$ is independent of $\tilde{W}$ and its probability density function is given by $p_{\tilde{X}(t_0)}$. 

The general system of discrete evolution equations described by Eq. (\ref{sde1}) can arise, for example, from stochastic differential equations \cite{fokker}. In particular, in many biological applications, solutions of linear stochastic differential equations are good fits to experimental single-molecule trajectories \cite{fokker}. As an example, we assume that the motion of the object of interest, e.g., a single molecule, is described by the following linear vector stochastic differential equation \cite{calderon3}
\begin{align}
\label{lineareq}
dX(\tau)=\left(V+F(\tau)X(\tau)\right)d\tau+G(\tau)dB(\tau),\quad\quad\tau\geq t_0,
\end{align}  
where the 3D random process $X(\tau)$ describes the location of the object at time $\tau\geq t_0$, $F\in\mathbb{R}^{3\times 3}$ and $G\in\mathbb{R}^{3\times r}$ are continuous matrix time-functions related to the first order drift and diffusion coefficients, respectively, $V\in\mathbb{R}^3$ is the zero order drift coefficient, and $\left\{B(\tau)\in\mathbb{R}^r,\right.$ $\left.\tau\geq t_0\right\}$ is a random process \cite{basharov}.

Here, we assume that $\left\{B(\tau)\in\mathbb{R}^r, \tau\geq t_0\right\}$ is an $r$-vector Brownian motion process with $E\left\{dB(\tau)dB(\tau)^T\right\}=I_{r\times r}, \tau\geq t_0$, where $I_{r\times r}$ is the $r\times r$ identity matrix \cite{calderon1,calderon2,calderon3}. Then, the solution of Eq. (\ref{lineareq}) at discrete time points $\tau_0:=t_0\leq\tau_1<\cdots<\tau_{l+1}<\cdots$ is given by \cite{jazwinski}
\begin{align}
\label{sdeConstant}
X(\tau_{l+1})=\phi(\tau_l,\tau_{l+1})X(\tau_l)+a(\tau_l,\tau_{l+1})+W_g(\tau_l,\tau_{l+1}),
\end{align}
where the continuous matrix time-function $\phi\in\mathbb{R}^{3\times 3}$ is given by
\begin{align*}
&\frac{d\phi(t,\tau)}{dt}=F(t)\phi(t,\tau),\quad \phi(\tau,\tau)=I_{3\times 3},\quad \mbox{for all}\ t,\tau\geq t_0,\\
&\phi(t,\tau)\phi(\tau,\psi)=\phi(t,\psi),\quad \mbox{for all}\ t, \tau, \psi\geq t_0,
\end{align*}
and the vector $a(\tau_l,\tau_{l+1})\in\mathbb{R}^{3\times 1}$ is given by
\begin{align*} 
a(\tau_l,\tau_{l+1}):=\int_{\tau_l}^{\tau_{l+1}}\phi(\tau,\tau_{l+1})Vd\tau.
\end{align*}
Also, in this case, $\left\{W_g(\tau_l,\tau_{l+1}):=\int_{\tau_l}^{\tau_{l+1}}\phi(\tau,\tau_{l+1})G(\tau)dB(\tau), l=1,2,\cdots\right\}$ is a zero mean white Gaussian sequence with covariance $Q_g(\tau_l,\tau_{l+1})\in\mathbb{R}^{3\times 3}$ given by
\begin{align*}
Q_g(\tau_l,\tau_{l+1})=\int_{\tau_l}^{\tau_{l+1}}\phi(\tau,\tau_{l+1})G(\tau)G^T(\tau)\phi^T(\tau,\tau_{l+1})d\tau.
\end{align*}
By letting $X(\tau)=H\tilde{X}(\tau)=I_{3\times 3}\tilde{X}(\tau)=\tilde{X}(\tau),\tau\geq t_0$, and $\phi(\tau_l,\tau_{l+1})=\tilde{\phi}(\tau_l,\tau_{l+1}),$ we obtain expressions of the form of Eq. (\ref{sde1}), where we assume that
\begin{align*} 
\left\{\tilde{W}(\tau_l,\tau_{l+1})=a(\tau_l,\tau_{l+1})+W_g(\tau_l,\tau_{l+1}), l=1,2,\cdots\right\}
\end{align*}
is a white Gaussian sequence with mean $a(\tau_l,\tau_{l+1})$ and covariance $Q_g(\tau_l,\tau_{l+1})$.

As an another example, for pure diffusion motion, when $V$ and $F(\tau), \tau\geq 0$, in Eq. (\ref{lineareq}) are equal to zero, the discrete motion model is given by   
\begin{align}
X(\tau_{l+1})=X(\tau_l)+W_g(\tau_l,\tau_{l+1}), \quad \tau_0:=t_0\leq\tau_1<\cdots<\tau_{l+1}<\cdots.
\end{align}
Setting $\tilde{X}(\tau):=X(\tau), \tau\geq t_0$, with $H$ the identity matrix, $\tilde{\phi}(\tau_l,\tau_{l+1})=\phi(\tau_l,\tau_{l+1})=I_{3\times 3}$, and $\tilde{W}(\tau_l,\tau_{l+1})=W_g(\tau_l,\tau_{l+1})$, we again obtain expressions of the form of Eq. (\ref{sde1}).

The above discussion motivates us to model the motion of the object, in the following definition, by Eq. (\ref{sde1}) with, in general, an arbitrary distributed process noise $\tilde{W}$. In particular, we also consider the special case of Gaussian distributed process noise $\tilde{W}_g$, separately.

\begin{definition}
\label{def3}
Let $\mathcal{G}_{[t]}\Big(X,\left(\mathcal{U}_{[t]},\mathcal{T}_{[t]}\right),q,\mathcal{C},\Theta\Big)$ and $\mathcal{G}_L\Big(X,\left(\mathcal{U}_L,\mathcal{T}_L\right),q,\mathcal{C},\Theta\Big)$ be image detection processes driven by a stochastic trajectory $X$ and image function $q$ for a fixed time interval $[t_0,t]$ and for a fixed number $L$ of photons. Let $p_{X(t_0)}$ be the probability density function of the initial location $X(t_0)$ of the object. We assume that

a. the motion of the object is modeled through a general state space system with state $\tilde{X}(\tau)\in\mathbb{R}^k, \tau\geq t_0$, as 
\begin{align}
\label{systemEq_main}
\tilde{X}(\tau_{l+1})=\tilde{\phi}(\tau_l,\tau_{l+1})\tilde{X}(\tau_l)+\tilde{W}(\tau_l,\tau_{l+1}),\quad \tau_0:=t_0\leq\tau_1<\cdots<\tau_{l+1}<\cdots,
\end{align}
where we assume that there exists a matrix $H\in\mathbb{R}^{3\times k}$ such that $X(\tau)=H\tilde{X}(\tau), \tau\geq t_0$, $\tilde{\phi}(\tau_l,\tau_{l+1})\in\tilde{\Phi}$, where $\tilde{\Phi}=\left\{\tilde{\phi}(\tau,\psi)\right\}_{\psi>\tau\geq t_0}$ is a family of $k\times k$ invertible real-valued state-transition matrices, and $\left\{\tilde{W}(\tau_l,\tau_{l+1}),l=0,1,2,\cdots\right\}$ is a process noise sequence of independent $k$-dimensional random variables with probability density functions $p_{\tilde{W}(\tau_l,\tau_{l+1})}$.

b. We assume that
\begin{align}
\label{observeEq}
U_l=Z\left(X(\tau_l)\right),\quad l=1,2,\cdots,
\end{align}
where $\left\{Z\left(X(\tau_l)\right),l=1,2,\cdots\right\}$ is a measurement sequence of independent 2D random variables with probability density functions $p_{Z\left(X(\tau_l)\right)}=f_{X(\tau_l)}$, where $Z$ is a random function that maps the object space into the image space, $f_{X(\tau_l)}$ is the image profile of an object located at $X(\tau_l)$ defined in Definition \ref{def2} and 

c. We assume that the sequences $\left\{\tilde{W}(\tau_l,\tau_{l+1}), l=0,1,\cdots\right\}$, $\left\{Z\left(X(\tau_l)\right),l=1,2,\right.$ $\left.
\cdots\right\}$, and $\tilde{X}(t_0)$ are independent of one another.

The image detection process $\mathcal{G}_{[t]}\Big(X,\left(\mathcal{U}_{[t]},\mathcal{T}_{[t]}\right),q,\mathcal{C},\Theta\Big)$ (or $\mathcal{G}_L\Big(X,\left(\mathcal{U}_L,\mathcal{T}_L\right),q,\mathcal{C},\\
\Theta\Big)$) with the additional properties (a)-(c) is called an \emph{image detection process with expanded state space $\tilde{X}$ for a time interval $[t_0,t]$} (or for a fixed number $L$ of photons), and is denoted by $\mathcal{G}_{[t]}\Big(\left(\tilde{X},H,\tilde{W},Z\right),\left(\mathcal{U}_{[t]},\mathcal{T}_{[t]}\right),\tilde{\Phi},\mathcal{C},\Theta\Big)$ (or $\mathcal{G}_L\Big(\left(\tilde{X},H,\tilde{W},Z\right),\\
\left(\mathcal{U}_L,\mathcal{T}_L\right),\tilde{\Phi},\mathcal{C},\Theta\Big)$).

We further assume that 

$\alpha$. $\left\{\tilde{W}_g(\tau_l,\tau_{l+1}):=\tilde{W}(\tau_l,\tau_{l+1}), l=0,1,\cdots\right\}$ is a white Gaussian sequence with mean $\tilde{a}(\tau_l,\tau_{l+1})\in\mathbb{R}^k$ and covariance matrix $\tilde{Q}_g(\tau_l,\tau_{l+1})\in\mathbb{R}^{k\times k},\tilde{Q}_g(\tau_l,\tau_{l+1})>0$,

$\beta$. 
\begin{align}
\label{observeEq_gaussian}
Z(X(\tau_l)) = M'X(\tau_l)+Z_{g,l}\quad l=1,2,\cdots,
\end{align}
where $M':=\begin{bmatrix}M & 0_{2\times 1}\end{bmatrix}\in\mathbb{R}^{2\times 3}$, in which $M\in\mathbb{R}^{2\times 2}$ is an invertible magnification matrix used in the definition of the image function (Eq. (\ref{imagefunctionmain})), where $0_{2\times 1}$ is the $2\times 1$ zero matrix, and $\left\{Z_{g,l},l=1,2,\cdots\right\}$ is a measurement noise sequence of independent 2D Gaussian random variables with mean zero and the same covariance matrix $\Sigma_g\in\mathbb{R}^{2\times 2}, \Sigma_g>0$. 

$\gamma$. We assume that the initial state $\tilde{X}(t_0)$ is Gaussian distributed with mean $\bar{x}_0\in\mathbb{R}^k$ and covariance matrix $\tilde{P}_0\in\mathbb{R}^{k\times k}, \tilde{P}_0>0$.

If, in addition, an image detection process with expanded state space has the properties ($\alpha$)-($\gamma$), it is called an \emph{image detection process with expanded state space $\tilde{X}$ and Gaussian process and measurement noise models}, and is denoted by $\mathcal{G}^g_{[t]}\Big(\left(\tilde{X},H,\tilde{W}_g,\right.\Big.$ $\Big.\left.Z_g\right),\left(\mathcal{U}_{[t]},\mathcal{T}_{[t]}\right),\tilde{\Phi},M',\mathcal{C},\Theta\Big)$ (or $\mathcal{G}^g_L\Big(\left(\tilde{X},H,\tilde{W}_g,Z_g\right),\left(\mathcal{U}_L,\mathcal{T}_L\right),\tilde{\Phi},M',\mathcal{C},\Theta\Big)$) for a time interval $[t_0,t]$ (or for a fixed number $L$ of photons).
\end{definition}

In Corollary \ref{corollary1}, we calculated the probability density function of the image detection process in terms of the image function $q$ and the distribution $p_{pr_l}, l=1,2,\cdots$, of the prediction of the object location, given the previous locations of the detected photons and the current and previous time points. In the following theorem, for a linear stochastic system and Gaussian process and measurement noise, we calculate these distributions using the Kalman filter formulae. Also, for a more general Markov motion model described by a first order system with arbitrary distributed process and measurement noise, we calculate these distributions recursively. 

\begin{theorem}
\label{nonGaussian}
Let $\mathcal{G}_{[t]}\Big(\left(\tilde{X},H,\tilde{W},Z\right),\left(\mathcal{U}_{[t]},\mathcal{T}_{[t]}\right),\tilde{\Phi},\mathcal{C},\Theta\Big)$ (or $\mathcal{G}_L\Big(\left(\tilde{X},H,\tilde{W},\right.$ $\left.Z\right),\left(\mathcal{U}_L,\mathcal{T}_L\right),\tilde{\Phi},\mathcal{C},\Theta\Big)$) be an image detection process with expanded state space $\tilde{X}$ for a time interval $[t_0,t]$ (or for a fixed number $L$ of photons). Let $\mathcal{D}_k:=\left(\mathcal{U}_k,\mathcal{T}_k\right),\\
 k=0,1,\cdots$, and
\begin{align*}
p_{pr_l}\Big(x|\tau_l,d_{l-1}\Big):=p_{X(T_l)|T_l,\mathcal{D}_{l-1}}\Big(x|\tau_l,d_{l-1}\Big),\quad x\in\mathbb{R}^3,
\end{align*}
where $d_l\in\mathcal{C}^l\times\mathbb{R}^l_{[t]}$ (or $d_l\in\mathcal{C}^l\times\mathbb{R}^l_{[\infty]}$), be the probability density function of the prediction of the object location, and $p_{pr_1}\Big(x|\tau_1,d_0\Big):=p_{pr_1}\Big(x|\tau_1\Big)$.

1. Assume that there exist non-singular matrix $H_1\in\mathbb{R}^{3\times 3}$ and matrix $H_2\in\mathbb{R}^{3\times (k-3)}$ such that $H=\begin{bmatrix}H_1 & H_2\end{bmatrix}$. Let
\begin{align*}
S:=
\begin{bmatrix}
H_1 & H_2\\
0_{(k-3)\times 3} & I_{(k-3)\times (k-3)}
\end{bmatrix}\in\mathbb{R}^{k\times k}.
\end{align*}
Then, for $x:=\left(x_1,x_2,x_3\right)^T\in\mathbb{R}^3$ and $\bar{x}:=\left(x_1,x_2,x_3,\bar{x}_4,\cdots,\bar{x}_k\right)^T\in\mathbb{R}^k$, 
\begin{align*}
p_{pr_l}\Big(x|\tau_l,d_{l-1}\Big)=\int_{\mathbb{R}^{k-3}}\tilde{p}_{pr_l}\Big(S^{-1}\bar{x}|\tau_l,d_{l-1}\Big)\left|H_1\right|^{-1}d\bar{x}_4\cdots d\bar{x}_k,
\end{align*}
where $\tilde{p}_{pr_l}:=p_{\tilde{X}(T_l)|T_l,\mathcal{D}_{l-1}}, l=0,1,2,\cdots$, and $S^{-1}$ is given by 
\begin{align*}
S^{-1}=
\begin{bmatrix}
H^{-1}_1 & -H^{-1}_1H_2\\
0_{(k-3)\times 3} & I_{(k-3)\times (k-3)}
\end{bmatrix}.
\end{align*}
If $H=\begin{bmatrix}I_{3\times 3} & 0_{3\times (k-3)}\end{bmatrix}$, then,
\begin{align*}
p_{pr_l}\Big(x|\tau_l,d_{l-1}\Big)=\int_{\mathbb{R}^{k-3}}\tilde{p}_{pr_l}\Big(\bar{x}|\tau_l,d_{l-1}\Big)d\bar{x}_4\cdots d\bar{x}_k.
\end{align*}

2. The probability density function $\tilde{p}_{pr_l}, l=0,1,2,\cdots$, can be calculated through the following recursive formula, for $\bar{x}\in\mathbb{R}^k$,
\begin{small}
\begin{align}
\label{lemmaeq11}
\tilde{p}_{pr_{l+1}}\Big(\bar{x}|\tau_{l+1},d_l\Big)=\frac{1}{\left|\det\left(\phi(\tau_l,\tau_{l+1})\right)\right|}\int_{\mathbb{R}^k}\tilde{p}_{fi_l}\Big(\tilde{\phi}^{-1}(\tau_l,\tau_{l+1})\bar{x}_o|d_l\Big)p_{\tilde{W}(\tau_l,\tau_{l+1})}\Big(\bar{x}-\bar{x}_o\Big)d\bar{x}_o,
\end{align}
\end{small}
where $d_0=\emptyset$, and the distribution $\tilde{p}_{fi_l}\Big(\bar{x}|d_l\Big):=p_{\tilde{X}(T_l)|\mathcal{D}_l}\Big(\bar{x}|d_l\Big)$ of the filtered object location is given by
\begin{small}
\begin{align}
\label{lemmaeq22}
\tilde{p}_{fi_l}\Big(\bar{x}|d_l\Big)=\frac{p_{Z(H\bar{x})}\left(r_l\right)\tilde{p}_{pr_l}\Big(\bar{x}|\tau_l,d_{l-1}\Big)}{\int_{\mathbb{R}^k}p_{Z(H\bar{x}_o)}\left(r_l\right)\tilde{p}_{pr_l}\Big(\bar{x}_o|\tau_l,d_{l-1}\Big)d\bar{x}_o}.
\end{align}
\end{small}

3.1. Let $\mathcal{G}^g_{[t]}\Big(\left(\tilde{X},H,\tilde{W}_g,Z_g\right),\left(\mathcal{U}_{[t]},\mathcal{T}_{[t]}\right),\tilde{\Phi},M',\mathcal{C},\Theta\Big)$ (or $\mathcal{G}^g_L\Big(\left(\tilde{X},H,\tilde{W}_g,Z_g\right),\\
\left(\mathcal{U}_L,\mathcal{T}_L\right)\Big.$ $,\Big.\tilde{\Phi},M',\mathcal{C},\Theta\Big)$) be an image detection process with expanded state space $\tilde{X}$ and Gaussian process and measurement noise models for a time interval $[t_0,t]$ (or for a fixed number $L$ of photons). Let $C:=M'H$. Then, for $l=0,1,\cdots$, and $\bar{x}\in\mathbb{R}^k$,
\begin{small}
\begin{align}
\label{predictgauss}
\tilde{p}_{pr_{l+1}}\Big(\bar{x}|d_l,\tau_{l+1}\Big)=\frac{1}{(2\pi)^{k/2}\left[\det(P_{l+1}^l)\right]^{1/2}}\exp\left(-\frac{1}{2}(\bar{x}-\hat{x}_{l+1}^l)^T\left(P_{l+1}^l\right)^{-1}(\bar{x}-\hat{x}_{l+1}^l)\right),
\end{align}
\end{small}
where $d_l\in\mathcal{C}^l\times\mathbb{R}^l_{[t]}$ (or $d_l\in\mathcal{C}^l\times\mathbb{R}^l_{[\infty]}$), $\hat{x}_1^0=\tilde{\phi}(\tau_0,\tau_1)\bar{x}_0, P_1^0=\tilde{\phi}(\tau_0,\tau_1)\tilde{P}_0\tilde{\phi}^T(\tau_0,\tau_1)+\tilde{Q}_g(\tau_0,\tau_1)$, and for $l=1,2,\cdots$,
\begin{small}
\begin{align}
\label{kalman1}
&\hat{x}_{l+1}^l=\tilde{\phi}(\tau_l,\tau_{l+1})\hat{x}_l^l+\tilde{a}(\tau_l,\tau_{l+1}),\nonumber\\
&P_{l+1}^l=\tilde{\phi}(\tau_l,\tau_{l+1})P_l^l\tilde{\phi}^T(\tau_l,\tau_{l+1})+\tilde{Q}_g(\tau_l,\tau_{l+1}),
\end{align}
\end{small}
with
\begin{small}
\begin{align}
\label{kalman2}
&K_l=P_l^{l-1}C^T\left(CP_l^{l-1}C^T+\Sigma_g\right)^{-1},\nonumber\\
&\hat{x}_l^l=\hat{x}_l^{l-1}+K_l(r_l-C\hat{x}_l^{l-1}),\nonumber\\
&P_l^l=P_l^{l-1}-K_lCP_l^{l-1}.
\end{align}
\end{small}

3.2. Moreover, the conditional probability density function $p_{U_l|T_l,\mathcal{D}_{l-1}}$ is given by
\begin{small}
\begin{align}
\label{finalgauss}
p_{U_l|T_l,\mathcal{D}_{l-1}}\Big(r_l|\tau_l,d_{l-1}\Big)=\frac{1}{2\pi\left[\det\left(R_l\right)\right]^{1/2}}\exp\Bigg(-\frac{1}{2}(r_l-\hat{r}_l)^TR_l^{-1}(r_l-\hat{r}_l)\Bigg),
\end{align}
\end{small}
where $d_l\in\mathcal{C}^l\times\mathbb{R}^l_{[t]}$ (or $d_l\in\mathcal{C}^l\times\mathbb{R}^l_{[\infty]}$), $R_l:=CP_l^{l-1}C^T+\Sigma_g$ and $\hat{r}_l:=C\hat{x}_l^{l-1}$.
\end{theorem}

\begin{proof}
See Section \ref{proof:nonGaussian} in the supplementary material.
\end{proof}

\section{Maximum likelihood estimation}
\label{mle}

The main purpose of the presented materials in the previous section is to provide a mathematical framework to estimate the parameters of interest, such as the parameters of the model that describes the motion of a moving object with stochastic trajectories, from the acquired data. In this paper, we use the maximum likelihood estimation approach as follows. For a general parameter estimation problem, denoting the acquired data by $\bar{d}\in\mathbb{R}^m, m=1,2,\cdots$, the maximum likelihood estimate $\hat{\theta}_{mle}$ of $\theta\in\Theta$, if it exists, is given by
\begin{align*}
\hat{\theta}_{mle}=\argmin_{\theta\in\Theta}\Big(-\log\mathcal{L}(\theta|\bar{d})\Big),
\end{align*}
where $\mathcal{L}$ denotes the likelihood function. In our specific problem, the acquired data for the fixed time interval $[t_0,t]$ acquisition case is denoted by $\bar{d}_K\in\mathcal{C}^K\times\mathbb{R}_{[t]}^K, K=0,1,\cdots$. Then, the likelihood function $\mathcal{L}_{[t]}$ of $\mathcal{G}_{[t]}\Big(\left(\mathcal{U}_{[t]},\mathcal{T}_{[t]}\right),\mathcal{C},\Theta\Big)$ is given by, according to Theorem \ref{main1} (see also \cite{snyder,streit}), for $\theta\in\Theta$,
\begin{footnotesize}
\begin{align}
\label{likelihood_t}
\mathcal{L}_{[t]}(\theta|\bar{d}_K)=
\begin{cases}
e^{-\int_{t_0}^t\Lambda_{\theta}(\tau)d\tau}, & K=0,\\
e^{-\int_{t_0}^t\Lambda_{\theta}(\tau)d\tau}\prod_{k=1}^K\Lambda_{\theta}(\bar{\tau}_k)\left[\prod_{l=1}^Kp^{\theta}_{U_l|T_l,\mathcal{D}_{l-1}}\Big(\bar{r}_l|\bar{\tau}_l,\bar{d}_{l-1}\Big)\right], & K=1,2,\cdots,
\end{cases}
\end{align}
\end{footnotesize}
and the likelihood function $\mathcal{L}_L$ of $\mathcal{G}_L\Big(\left(\mathcal{U}_L,\mathcal{T}_L\right),\mathcal{C},\Theta\Big)$ is given by
\begin{align}
\label{likelihood_L}
\mathcal{L}_L(\theta|\bar{d}_L)=p^{\theta}_L(\bar{d}_L)=e^{-\int_{t_0}^{\bar{\tau}_L}\Lambda_{\theta}(\tau)d\tau}\prod_{k=1}^L\Lambda_{\theta}(\bar{\tau}_k)\left[\prod_{l=1}^Lp^{\theta}_{U_l|T_l,\mathcal{D}_{l-1}}\Big(\bar{r}_l|\bar{\tau}_l,\bar{d}_{l-1}\Big)\right],
\end{align}
where $\bar{d}_L\in\mathcal{C}^L\times\mathbb{R}_{[\infty]}^L, L=1,2,\cdots$.

In supplementary Section \ref{sup:ex1}, we provide an example to illustrate our results for the specific case that the motion model is described by a linear stochastic differential equation.

In the following, we present and discuss the results of the proposed maximum likelihood estimation method when applied to simulated data sets of trajectories of a single molecule.
 
\subsection{Simulated parameters} 
\label{parameters}

To analyze the performance of the proposed maximum likelihood estimation method, we simulated different data sets using parameters commonly used in single molecule experiments. Unless otherwise stated, the images of in-focus and out-of-focus molecules were generated with Airy and Born and Wolf profiles (Eqs. (\ref{airyprofileeq}) and (\ref{bornimagefunction})), respectively, where $n_a=1.4, \lambda=520$ nm, $n_o=1.515$, and $z_0=1$ $\mu$m. For the Gaussian measurement case, the image of a molecule was generated with a zero-mean Gaussian measurement noise with the probability density function given by Eq. (\ref{gaussianimagefunction}), where $\sigma=70$ nm, which is related to the corresponding Airy profile.

Furthermore, a measurement (magnification) matrix $M = 100I_{2\times 2}$ was assumed to map the object space to the image space.

\begin{figure}[htbp]
\centering\includegraphics[width=1\textwidth]{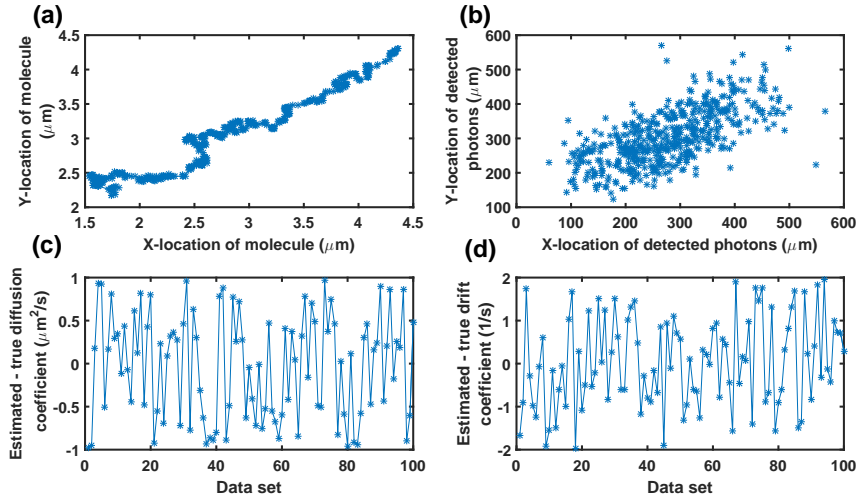}
\DeclareGraphicsExtensions{.eps} \caption{Analysis of the error of diffusion coefficient and first order drift coefficient estimates produced by the maximum likelihood estimation method for the Born and Wolf measurement model. (a) A trajectory of an out-of-focus molecule, with the out-of-focus level $z_0=1$ $\mu$m, in the object space simulated using Eq. (\ref{linearspecific}) where the time points are drawn from a Poisson process with mean 500 in the time interval $[0, 100]$ ms with the first order drift coefficient $F=-10I_{2\times 2}$/s and the diffusion coefficient $D=1$ $\mu\mbox{m}^2$/s. We assume the zero order drift is equal to 0. Also, we assume that the initial location of the molecule is Gaussian distributed with mean $x_0=(4.4,4.4)^T$ $\mu$m and covariance $P_0=10I_{2\times 2}$ n$\mbox{m}^2$. (b) Detected locations of the photons emitted from the molecule trajectory of part (a) in the image space which are simulated using Eq. (\ref{observeEq}) with the Born and Wolf profile (Eq. (\ref{bornimagefunction})) and the parameters given in Section \ref{parameters}. (c) Differences between the diffusion coefficient estimates and the true diffusion coefficient value for 100 data sets, each containing a trajectory of a molecule simulated using Eqs. (\ref{linearspecific}) and (\ref{observeEq}) with the Born and Wolf profile, and the parameters given in parts (a) and (b). (d) Differences between the first order drift coefficient estimates and its true value for the data sets of part (c).}
\label{fiveimage}
\end{figure}

\subsection{Estimation results}

\begin{figure}[htbp]
\centering\includegraphics[width=1\textwidth]{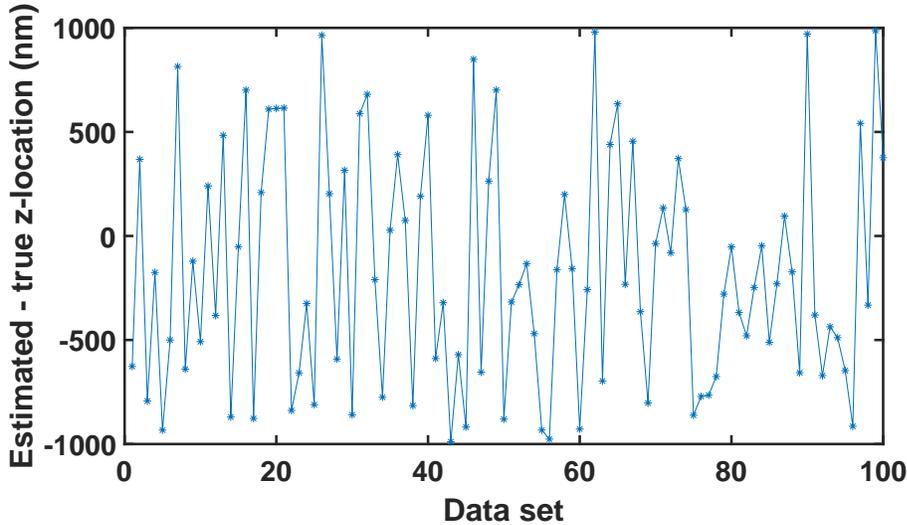}
\DeclareGraphicsExtensions{.eps} \caption{Analysis of the error of out-of-focus $z_0$-location estimates produced by the maximum likelihood estimation method for the Born and Wolf measurement model. Differences between the $z_0$-location estimates and its true value, $z_0=1$ $\mu$m, for the data sets of Fig. \ref{fiveimage}.}
\label{zlocation}
\end{figure}

Using simulated data sets, we first examine the performance of the maximum likelihood estimation method used to estimate the parameters of the linear motion model of a moving molecule in terms of the bias of the method. The bias is assessed by the average of the deviations of the estimates from the true value. For this purpose, we simulated 100 data sets, each containing a trajectory of an out-of-focus molecule, with the out-of-focus level $z_0=1$ $\mu$m, simulated using Eqs. (\ref{linearspecific}) and (\ref{observeEq}), with the Born and Wolf profile (Eq. (\ref{bornimagefunction})) and the parameters given in Section \ref{parameters}, with a mean photon count of 500 photons in the time interval $\left[0,100\right]$ ms, where the first order drift coefficient $F=-10I_{2\times 2}$/s and the diffusion coefficient $D=1$ $\mu\mbox{m}^2$/s. We assume the zero order drift is equal to 0. In Figs. \ref{fiveimage}(a) and \ref{fiveimage}(b), an example of a molecule trajectory in the object space and its image in the image space are shown. For these data sets, we calculated the maximum likelihood estimates of the diffusion and drift coefficients, separately. For this purpose, we needed to obtain the distributions of the prediction in the likelihood function expressions (Eqs. (\ref{likelihood_t}) and (\ref{likelihood_L})) through Eqs. (\ref{lemmaeq11}) and (\ref{lemmaeq22}), which in general is a computationally expensive problem. We approximated the distributions of the prediction using a sequential Monte Carlo algorithm proposed in \cite{automatica}. The overall approach is explained in supplementary Section \ref{basicparticlefilter} in detail. In Figs. \ref{fiveimage}(c) and \ref{fiveimage}(d), the differences between the maximum likelihood estimates of the diffusion and the first order drift coefficients and the true values are plotted. We also estimated the $z_0$-location of the molecule, i.e., the out-of-focus level, and show the errors of estimation in Fig. \ref{zlocation}. As can be seen, the deviations of the estimates from the ground truth are, overall, centered around 0 nm, which suggests that there is no systematic bias associated with our proposed method (the average of the diffusion coefficient deviations and the first order drift coefficient deviations are -0.0319 $\mu\mbox{m}^2$/s and 0.0307/s, respectively).

\begin{figure}[htbp]
\centering\includegraphics[width=1\textwidth]{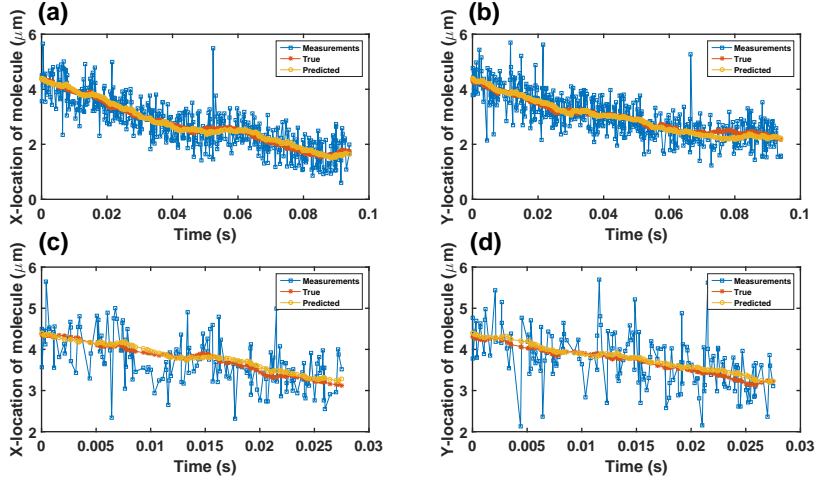}
\DeclareGraphicsExtensions{.eps} \caption{Predicted locations of the molecule for the Born and Wolf measurement model. (a) and (b) Means of the distributions of the prediction of the molecule $x$- and $y$-locations and the true $x$- and $y$-locations of the molecule for the same data set as in Figs. \ref{fiveimage}(a) and \ref{fiveimage}(b). The measurements transformed from the image space to the object space are also shown. (c) and (d) Means of the distributions of the prediction of the molecule $x$- and $y$-locations and the true $x$- and $y$-locations of the molecule over the time interval $[0,27.5]$ ms.}
\label{siximage}
\end{figure} 

We further investigate the distribution $p_{pr_l}, l=1,2,\cdots$, of the prediction of the molecule location, given previous observations, for the molecule trajectory shown in Figs. \ref{fiveimage}(a) and \ref{fiveimage}(b). The means of the distributions of the prediction of the molecule $x$- and $y$-locations and the true $x$- and $y$-locations are shown in Fig. \ref{siximage}(a) and \ref{siximage}(b). We also show the measurements transformed from the image space to the object space, which are obtained as follows. The location $X_o:=(x_o,y_o,z_o)^T\in\mathbb{R}^3$ in the object space is transformed into the location $X_i:=(x_i,y_i)^T\in\mathbb{R}^2$ in the image space through a linear map as, for $M'\in\mathbb{R}^{2\times 3}$,
\begin{equation}
X_i=M'X_o.
\end{equation}
In Fig. \ref{fiveimage}, where we have a trajectory of an out-of-focus molecule, with the out-of-focus plane $z_o = 1$ $\mu$m, it is assumed that the magnification matrix (measurement mapping matrix) $M'=100I_{2\times 2}$. Then, the $x$- and $y$-locations of the measurements mapped to the object space are obtained as
\begin{equation}
x_o=x_i/100,\quad y_o=y_i/100.
\end{equation}
For a better visual comparison, the means of the distributions of the prediction of the molecule locations and the true locations for $x$- and $y$-coordinates are also shown over a shorter time interval in Figs. \ref{siximage}(c) and \ref{siximage}(d). As can be seen, the predicted locations are able to track the true locations of the molecule for both $x$- and $y$-coordinates. We also show the differences between the means of the distributions of the prediction of the molecule locations and the true locations of the molecule in Fig. \ref{error_born} (see Section \ref{appendixfigures} in the supplementary material). We also applied the proposed method to trajectory data of an in-focus molecule simulated using an Airy profile, with the same standard deviation as the Born and Wolf data, and obtained similar results (see Figs. \ref{mle_estimates_airy}, \ref{predicted_airy} and \ref{error_airy} in supplementary Section \ref{airy_sup}).

\begin{figure}[htbp]
\centering\includegraphics[width=1\textwidth]{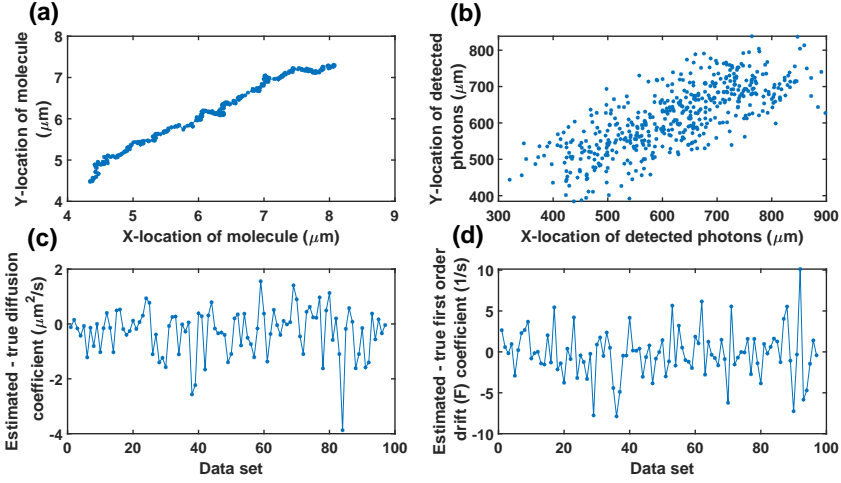}
\DeclareGraphicsExtensions{.eps} \caption{Analysis of the error of diffusion coefficient and first order drift coefficient estimates produced by the maximum likelihood estimation method for the Gaussian measurement noise case. (a) A 2D trajectory of an in-focus molecule in the object space simulated using Eq. (\ref{linearspecific}) where the time points are drawn from a Poisson process with mean 500 in the time interval $[0, 100]$ ms with the first order drift coefficient $F=-10I_{2\times 2}$/s and the diffusion coefficient $D=1$ $\mu\mbox{m}^2$/s. Also, we assume that the zero order drift coefficient $V=\left(100,100\right)^T$ $\mu$/s, the initial location of the molecule is Gaussian distributed with mean $x_0=(4.4,4.4)^T$ $\mu$m and covariance $P_0=10I_{2\times 2}$ n$\mbox{m}^2$. (b) Detected locations of the photons emitted from the molecule trajectory of part (a) in the image space which are simulated using Eq. (\ref{observeEq_gaussian}) with the Gaussian measurement noise (Eq. (\ref{gaussianimagefunction})) and $\sigma=0.51$ $\mu$m. (c) Differences between the diffusion coefficient estimates and the true diffusion coefficient value for 100 data sets, each containing a trajectory of a molecule simulated using Eqs. (\ref{linearspecific}) and (\ref{observeEq_gaussian}) with the Gaussian profile, and the parameters given in parts (a) and (b). (d) Differences between the first order drift coefficient estimates and its true value for the data sets of part (c).}
\label{gaussian_estimation}
\end{figure}

\begin{figure}[htbp]
\centering\includegraphics[width=1\textwidth]{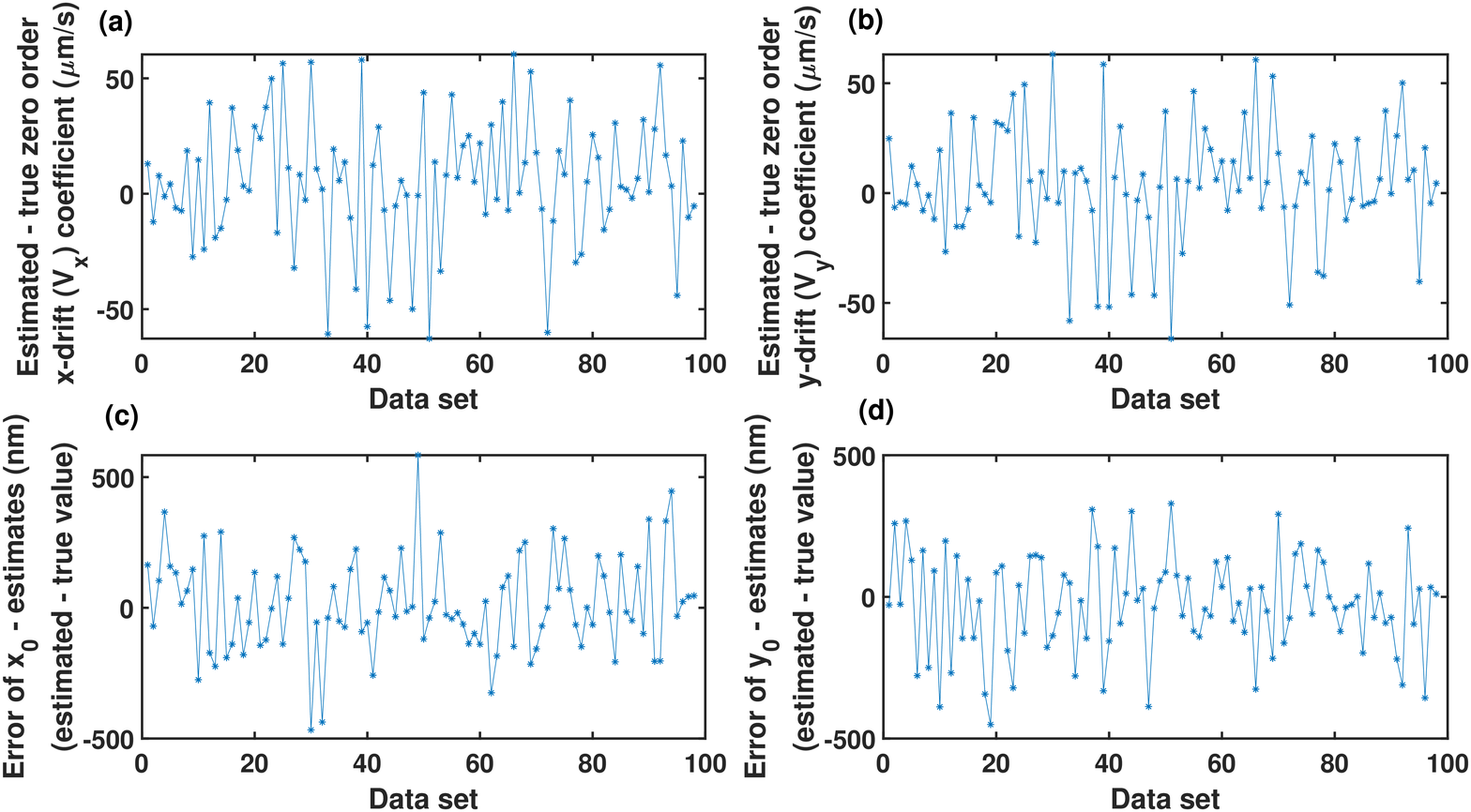}
\DeclareGraphicsExtensions{.eps} \caption{Analysis of the error of the zero order drift coefficient and initial location estimates produced by the maximum likelihood estimation method for the Gaussian measurement noise case. (a) and (b) Differences between the zero order drift coefficient estimates and its true value, in both $x$- and $y$-directions, for the data sets of Fig. \ref{gaussian_estimation}. (c) and (d) Differences between the initial location estimates and its true value, in both $x$- and $y$-directions, for the data sets of Fig. \ref{gaussian_estimation}.}
\label{gaussian_estimation2}
\end{figure}

\begin{figure}[htbp]
\centering\includegraphics[width=1\textwidth]{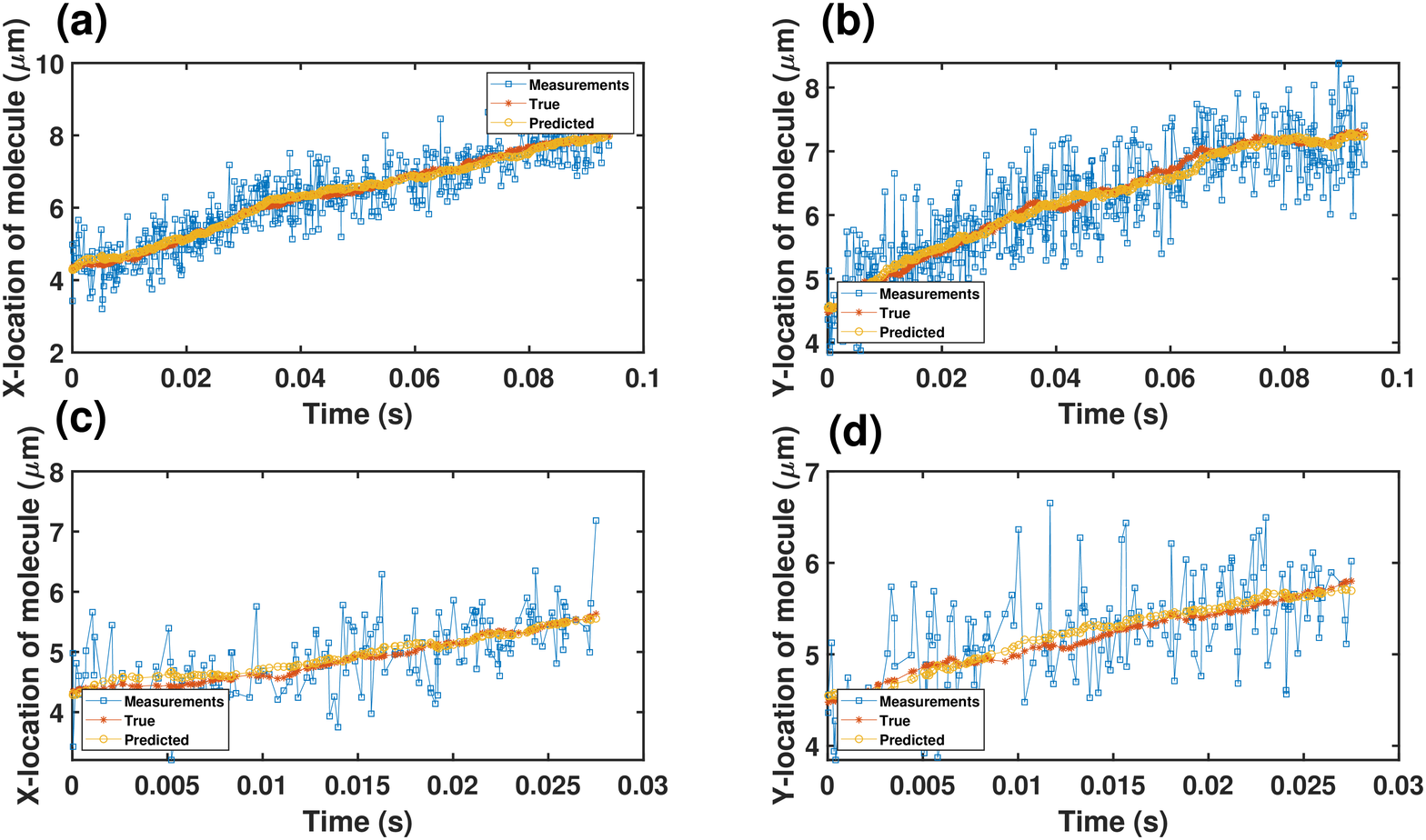}
\DeclareGraphicsExtensions{.eps} \caption{Predicted locations of the molecule for the Gaussian measurement noise case. (a) and (b) Means of the distributions of the prediction of the molecule $x$- and $y$-locations and the true $x$- and $y$-locations of the molecule for the same data set as in Figs. \ref{gaussian_estimation}(a) and \ref{gaussian_estimation}(b). The measurements transformed from the image space to the object space are also shown. (c) and (d) Means of the distributions of the prediction of the molecule $x$- and $y$-locations and the true $x$- and $y$-locations of the molecule over the time interval $[0,27.5]$ ms.}
\label{gaussian_prediction}
\end{figure}

As mentioned, in some applications, it is useful to approximate the point spread function of an optical system with a Gaussian profile. We analyzed the error of the estimates for simulated data sets with Gaussian measurement noise, with the same standard deviation as the Born and Wolf data, and obtained similar results (see Figs. \ref{gaussian_estimation}, \ref{gaussian_estimation2}, \ref{gaussian_prediction} and \ref{error_gaussian}). This time we estimated all the parameters of the trajectory together, i.e., we assumed that the parameter vector $\theta:=\left(V,F,D,x_0,y_0\right)$, where $V\in\mathbb{R}^2$ and $F\in\mathbb{R}$ denote the zero order and first order drift, respectively, $D\in\mathbb{R}$ is the diffusion coefficient and $\left(x_0,y_0\right)\in\mathbb{R}^2$ is the initial location of the molecule. We also consider the more general case where $F=\begin{bmatrix}F_x & 0 \\ 0 & F_y \end{bmatrix}, F_x,F_y\in\mathbb{R}$ (Fig. \ref{complex_drift}). In order to calculate the predicted locations of the molecule for Gaussian measurements, we took advantage of the relationship between the likelihood function and Kalman filter formulae (see Theorem \ref{nonGaussian}). It improved the computational efficiency significantly.

\section{Fisher information matrix and CRLB}
\label{crlbsection}

In any estimation problem, the performance of the estimator can be evaluated by calculating their standard deviations from the true parameter values. According to the Cram\'er-Rao inequality, the covariance matrix of any unbiased estimator $\hat{\theta}$ of an unknown vector parameter $\theta$ is bounded from below by the inverse of the Fisher information matrix $I(\theta)$, i.e., $\mbox{Cov}(\hat{\theta})\geq I^{-1}(\theta)$. Therefore, a benchmark on the standard deviation of estimates can be obtained by the square root of the inverse of the Fisher information matrix. Note that the Fisher information matrix only depends on the statistical nature of the acquired data and is independent of the applied estimation technique. Since this concept is very important when we have fixed time points, as we defined image detection processes and their probability density functions at fixed time points in Section \ref{fundamentalsection}, here, we first introduce a notation for the Fisher information matrix of these processes in Definition \ref{fish_fixed}, and use it to calculate the Fisher information matrix of image detection processes for the fixed time interval and for the fixed number of photons in Theorem \ref{main2}. 

\begin{definition}
\label{fish_fixed}
For $t_0\leq\tau_1<\cdots<\tau_K$, let $\mathcal{G}_{\tau_1,\cdots,\tau_K}\Big(\left(\mathcal{U}_K,\mathcal{T}_K\right),\mathcal{C},\Theta\Big)$ be an image detection process at fixed time points $\tau_1,\cdots,\tau_K$. We introduce the following notation for the Fisher information matrix of $\mathcal{G}_{\tau_1,\cdots,\tau_K}\Big(\left(\mathcal{U}_K,\mathcal{T}_K\right),\mathcal{C},\Theta\Big)$ as, for a row parameter vector $\theta\in\Theta$,
\begin{footnotesize}
\begin{align*}
I_{\tau_1,\cdots,\tau_K}(\theta):&=E_{\mathcal{U}_K|\mathcal{T}_K=\tau_{1:K}}\left\{\left(\frac{\partial \log p^{\theta}_{\mathcal{U}_K|\mathcal{T}_K}\Big(r_{1:K}|\tau_{1:K}\Big)}{\partial\theta}\right)^T\left(\frac{\partial \log p^{\theta}_{\mathcal{U}_K|\mathcal{T}_K}\Big(r_{1:K}|\tau_{1:K}\Big)}{\partial\theta}\right)\right\}\nonumber\\
&=\int_{\mathcal{C}}\cdots\int_{\mathcal{C}}p^{\theta}_{\mathcal{U}_K|\mathcal{T}_K}\Big(r_{1:K}|\tau_{1:K}\Big)\left(\frac{\partial \log p^{\theta}_{\mathcal{U}_K|\mathcal{T}_K}\Big(r_{1:K}|\tau_{1:K}\Big)}{\partial\theta}\right)^T\nonumber\\
&\ \ \ \ \ \ \ \ \ \ \ \ \ \ \ \ \ \ \ \ \ \ \ \ \times\left(\frac{\partial \log p^{\theta}_{\mathcal{U}_K|\mathcal{T}_K}\Big(r_{1:K}|\tau_{1:K}\Big)}{\partial\theta}\right)dr_1\cdots dr_K,
\end{align*}
\end{footnotesize}
for $t_0\leq\tau_1<\cdots<\tau_K$, and $I_{\tau_1,\cdots,\tau_K}(\theta)=0$, otherwise, where $r_{1:K}:=\left(r_1,\cdots,r_K\right)$\\
$,r_1,\cdots,r_k\in\mathcal{C}, \tau_{1:K}:=\left(\tau_1,\cdots,\tau_K\right), K=1,2,\cdots$, and $E_{\mathcal{U}_K|\mathcal{T}_K=\tau_{1:K}}$ denotes the expected value with respect to the conditional probability density function $p^{\theta}_{\mathcal{U}_K|\mathcal{T}_K}$ of $\mathcal{U}_K$, given $\mathcal{T}_K=\tau_{1:K}$.
\end{definition}

\begin{theorem}
\label{main2}
Let $\mathcal{G}_{[t]}\Big(\left(\mathcal{U}_{[t]},\mathcal{T}_{[t]}\right),\mathcal{C},\Theta\Big)$ and $\mathcal{G}_L\Big(\left(\mathcal{U}_L,\mathcal{T}_L\right),\mathcal{C},\Theta\Big)$ be image detection processes for a time interval $[t_0,t]$ and for a fixed number $L$ of photons, respectively. Let $\mathcal{D}_{[t]}:=\left(\mathcal{U}_{[t]},\mathcal{T}_{[t]}\right), \mathcal{D}_k:=\left(\mathcal{U}_k,\mathcal{T}_k\right), k=0,1,\cdots$. Assume that the conditional probability density functions $p^{\theta}_{U_l|T_l,\mathcal{D}_{l-1}}, l=1,2,\cdots$, of $U_l$, given $T_l$ and $\mathcal{D}_{l-1}$, satisfy the following regularity conditions, for $\theta=(\theta_1,\cdots,\theta_n)\in\Theta$,

(a) $\frac{\partial p^{\theta}_{U_l|T_l,\mathcal{D}_{l-1}}\Big(r_l|\tau_l,d_{l-1}\Big)}{\partial\theta_i}$ exists for $i=1,\cdots,n$,
 
(b) $\bigint_{\mathcal{C}}\left|\frac{\partial  p^{\theta}_{U_l|T_l,\mathcal{D}_{l-1}}\Big(r|\tau_l,d_{l-1}\Big)}{\partial\theta_i}\right|dr<\infty$ for $i=1,\cdots,n$,

where $d_l\in\mathcal{C}^l\times\mathbb{R}_{[t]}^l$ for $\mathcal{G}_{[t]}$, $d_l\in\mathcal{C}^l\times\mathbb{R}_{[\infty]}^l$ for $\mathcal{G}_{L}$, and $p^{\theta}\Big(r_1|\tau_1,d_0\Big):=p^{\theta}\Big(r_1|\tau_1\Big)$.

1.1. Then, the Fisher information matrix $I_{[t]}$ of $\mathcal{G}_{[t]}$ is given by
\begin{footnotesize}
\begin{align}
\label{theroem2_firsteq}
I_{[t]}(\theta)&=\frac{1}{P_{\theta}\Big(N(t)=0\Big)}\left(\frac{\partial P_{\theta}\Big(N(t)=0\Big)}{\partial\theta}\right)^T\left(\frac{\partial P_{\theta}\Big(N(t)=0\Big)}{\partial\theta}\right)\nonumber\\
&\ \ +\sum_{K=1}^{\infty}\int_{t_0}^t\int_{t_0}^{\tau_K}\cdots\int_{t_0}^{\tau_3}\int_{t_0}^{\tau_2}\Bigg[\int_{\mathcal{C}}\cdots\int_{\mathcal{C}}\frac{1}{p^{\theta}_{[t]}\Big(d_K,K\Big)}\left(\frac{\partial p^{\theta}_{[t]}\Big(d_K,K\Big)}{\partial\theta}\right)^T \left(\frac{\partial p^{\theta}_{[t]}\Big(d_K,K\Big)}{\partial\theta}\right)\Bigg.\nonumber\\
&\Bigg.\ \ \ \ \ \ \ \ \ \ \ \times dr_1\cdots dr_K\Bigg]d\tau_1d\tau_2\cdots d\tau_{K-1}d\tau_K,
\end{align}
\end{footnotesize}
where $d_l\in\mathcal{C}^l\times\mathbb{R}_{[t]}^l$, and $p^{\theta}_{[t]}$ denotes the probability density function of $\mathcal{D}_{[t]}$ and $N(t)$.

1.2. Assume that the photon detection rate $\Lambda$ is independent of $\theta$. Then, $I_{[t]}$ can be calculated as
\begin{align}
\label{fixed1_1}
I_{[t]}(\theta)&=e^{-\int_{t_0}^t\Lambda(\tau)d\tau}\sum_{K=1}^{\infty}\Bigg\{\int_{t_0}^t\int_{t_0}^{\tau_K}\cdots\int_{t_0}^{\tau_3}\int_{t_0}^{\tau_2}I_{\tau_1,\cdots,\tau_K}(\theta)\prod_{k=1}^K\Lambda(\tau_k)\Bigg.\nonumber\\
&\Bigg.\ \ \ \ \ \ \ \ \ \ \ \ \ \ \ \ \ \times d\tau_1d\tau_2\cdots d\tau_{K-1}d\tau_K\Bigg\},
\end{align}
where the Fisher information matrix $I_{\tau_1,\cdots,\tau_K}$ of the image detection process at fixed time points $\tau_1,\cdots,\tau_K$ $\mathcal{G}_{\tau_1,\cdots,\tau_K}\Big(\left(\mathcal{U}_L,\mathcal{T}_L\right),\mathcal{C},\Theta\Big)$ is given by
\begin{footnotesize}
\begin{align}
\label{fixed2}
I_{\tau_1,\cdots,\tau_K}(\theta)=
\begin{cases}
\sum_{l=1}^KI^{\tau_1,\cdots,\tau_l}_{U_l|T_l,\mathcal{D}_{l-1}}(\theta), & t_0\leq\tau_1<\cdots<\tau_K\leq t,\\
0, & \mbox{otherwise},
\end{cases}
\end{align}
\end{footnotesize}
in which the Fisher information matrix $I^{\tau_1,\cdots,\tau_l}_{U_l|T_l,\mathcal{D}_{l-1}}$ calculated with respect to the conditional probability density function $p^{\theta}_{U_l|T_l,\mathcal{D}_{l-1}}$ at fixed time points $\mathcal{T}_l=\tau_{1:l}$ is given by
\begin{scriptsize}
\begin{align}
\label{fixed3}
I^{\tau_1,\cdots,\tau_l}_{U_l|T_l,\mathcal{D}_{l-1}}(\theta)&=E_{\mathcal{U}_l|\mathcal{T}_l=\tau_{1:l}}\left\{\left(\frac{\partial \log p^{\theta}_{U_l|T_l,\mathcal{D}_{l-1}}\Big(r_l|\tau_l,d_{l-1}\Big)}{\partial\theta}\right)^T\left(\frac{\partial \log p^{\theta}_{U_l|T_l,\mathcal{D}_{l-1}}\Big(r_l|\tau_l,d_{l-1}\Big)}{\partial\theta}\right)\right\}\nonumber\\
&=\int_{\mathcal{C}}\cdots\int_{\mathcal{C}}p^{\theta}_{\mathcal{U}_{l-1}|\mathcal{T}_{l-1}}\Big(r_{1:l-1}|\tau_{1:l-1}\Big)\Bigg[\int_{\mathcal{C}} \frac{1}{p^{\theta}_{U_l|T_l,\mathcal{D}_{l-1}}\Big(r_l|\tau_l,d_{l-1}\Big)}\Bigg.\nonumber\\
&\Bigg.\ \ \ \ \ \ \times\left(\frac{\partial p^{\theta}_{U_l|T_l,\mathcal{D}_{l-1}}\Big(r_l|\tau_l,d_{l-1}\Big)}{\partial\theta}\right)^T\left(\frac{\partial p^{\theta}_{U_l|T_l,\mathcal{D}_{l-1}}\Big(r_l|\tau_l,d_{l-1}\Big)}{\partial\theta}\right)dr_l\Bigg]dr_{l-1}\cdots dr_1,
\end{align}
\end{scriptsize}
with $r_{1:l}:=\left(r_1,\cdots,r_l\right), \tau_{1:l}:=\left(\tau_1,\cdots,\tau_l\right)$, and $I^{\tau_1}_{U_1|T_1}$ given by
\begin{small}
\begin{align}
\label{initialconndi}
I^{\tau_1}_{U_1|T_1}(\theta)=\int_{\mathcal{C}} \frac{1}{p^{\theta}_{U_1|T_1}\Big(r|\tau_1\Big)}\left(\frac{\partial p^{\theta}_{U_1|T_1}\Big(r|\tau_1\Big)}{\partial\theta}\right)^T\left(\frac{\partial p^{\theta}_{U_1|T_1}\Big(r|\tau_1\Big)}{\partial\theta}\right)dr.
\end{align}
\end{small}

2.1. The Fisher information matrix $I_L$ of $\mathcal{G}_L$ is given by
\begin{footnotesize}
\begin{align*}
I_L(\theta)&=\int_{t_0}^{\infty}\int_{t_0}^{\tau_L}\cdots\int_{t_0}^{\tau_3}\int_{t_0}^{\tau_2}\Bigg[\int_{\mathcal{C}}\cdots\int_{\mathcal{C}}\frac{1}{p^{\theta}_L\Big(d_L\Big)}\left(\frac{\partial p^{\theta}_L\Big(d_L\Big)}{\partial\theta}\right)^T \left(\frac{\partial p^{\theta}_L\Big(d_L\Big)}{\partial\theta}\right)dr_1\cdots dr_L\Bigg]\nonumber\\
&\ \ \ \ \ \ \ \ \ \ \ \ \ \ \ \ \ \times d\tau_1d\tau_2\cdots d\tau_{L-1}d\tau_L,
\end{align*}
\end{footnotesize}
where $d_l\in\mathcal{C}^l\times\mathbb{R}_{[\infty]}^l$, and $p^{\theta}_L$ denotes the probability density function of $\mathcal{D}_L$.

2.2. Assume that the photon detection rate $\Lambda$ is independent of $\theta$. Then, $I_L$ can be obtained as
\begin{align}
\label{fishi2_2}
I_L(\theta)&=\int_{t_0}^{\infty}\int_{t_0}^{\tau_L}\cdots\int_{t_0}^{\tau_3}\int_{t_0}^{\tau_2}I_{\tau_1,\cdots,\tau_L}(\theta)e^{-\int_{t_0}^{\tau_L}\Lambda(\tau)d\tau}\prod_{k=1}^L\Lambda(\tau_k)\nonumber\\
&\ \ \ \ \ \ \ \ \ \ \ \ \ \ \ \ \ \times d\tau_1d\tau_2\cdots d\tau_{L-1}d\tau_L.
\end{align}
\end{theorem}

\begin{remark}
Note that for $K=1$, the time integral of Eq. (\ref{fixed1_1}) is calculated over the interval $[t_0,t]$, i.e., $\int_{t_0}^tI_{\tau_1}(\theta)\Lambda(\tau_1)d\tau_1$.
\end{remark}

\begin{proof}
See Section \ref{proof:main2} in the supplementary material.
\end{proof}

We next derive expressions for the Fisher information matrices of the image detection processes driven by the stochastic trajectory $X$ and image function $q$ for a time interval $[t_0,t]$ and for a fixed number $L$ of photons in the following corollary to Theorem \ref{main2}.

\begin{corollary}
\label{corollary2}
Let $\mathcal{G}_{[t]}\Big(X,\left(\mathcal{U}_{[t]},\mathcal{T}_{[t]}\right),q,\mathcal{C},\Theta\Big)$ (or $\mathcal{G}_L\Big(X,\left(\mathcal{U}_L,\mathcal{T}_L\right),q,\mathcal{C},\Theta\Big)$) be an image detection process driven by the stochastic trajectory $X$ and image function $q$ for a time interval $[t_0,t]$ (or for a fixed number $L$ of photons). Let, for a row parameter vector $\theta=(\theta_1,\cdots,\theta_n)\in\Theta$, the $n$-dimensional vector $F^{\theta}_l$ be given by
\begin{align}
\label{q2}
F^{\theta}_l\Big(x,d_l\Big):=\underbrace{\begin{bmatrix}\left(df^{\theta}_x\left(r_l\right)\right)^T &  \left(dp_{pr_l}^{\theta}\Big(x|\tau_l,d_{l-1}\Big)\right)^T\end{bmatrix}}_{\mbox{Block row vector}}\begin{bmatrix}p_{pr_l}^{\theta}\Big(x|\tau_l,d_{l-1}\Big) \\ f_x^{\theta}\left(r_l\right)\end{bmatrix},\quad x\in\mathbb{R}^3,
\end{align}
where $d_l\in\mathcal{C}^l\times\mathbb{R}^l_{[t]}$ (or $d_l\in\mathcal{C}^l\times\mathbb{R}^l_{[\infty]}$), $r_{1:l}:=\left(r_1,\cdots,r_l\right), \tau_{1:l}:=\left(\tau_1,\cdots,\tau_l\right)$, $p_{pr_l}^{\theta}:=p^{\theta}_{X(T_l)|T_l,\mathcal{D}_{l-1}}, p_{pr_1}^{\theta}\Big(x|\tau_1,d_0\Big):=p_{pr_1}^{\theta}\Big(x|\tau_1\Big)$, denotes the distribution of the prediction of the object location, and $dp_{pr_l}^{\theta}:=\frac{\partial p_{pr_l}^{\theta}}{\partial\theta}, df_x^{\theta}:=\frac{\partial f_x^{\theta}}{\partial\theta}$. Assume that the photon detection rate $\Lambda$ is independent of $\theta$. Then, $I_{\tau_1,\cdots,\tau_K}$ in Eq. (\ref{fixed1_1}) (or Eq. (\ref{fishi2_2})) of Theorem \ref{main2} is given by
\begin{align*}
I_{\tau_1,\cdots,\tau_K}(\theta)=
\begin{cases}
\sum_{l=1}^KI^{\tau_1,\cdots,\tau_l}_{U_l|T_l,\mathcal{D}_{l-1}}(\theta), & t_0\leq\tau_1<\cdots<\tau_K\leq t,\\
0, & \mbox{otherwise},
\end{cases}
\end{align*}
where 
\begin{footnotesize}
\begin{align}
\label{fixed3_2}
I^{\tau_1,\cdots,\tau_l}_{U_l|T_l,\mathcal{D}_{l-1}}(\theta)&=\int_{\mathcal{C}}\cdots\int_{\mathcal{C}}p^{\theta}_{\mathcal{U}_{l-1}|\mathcal{T}_{l-1}}\Big(r_{1:l-1}|\tau_{1:l-1}\Big)\nonumber\\
&\ \ \ \ \ \ \ \ \ \ \ \ \times\left[\int_{\mathbb{R}^3}\int_{\mathbb{R}^3}\left(\int_{\mathcal{C}}\frac{F^{\theta}_l\Big(x_1,d_l\Big)\left[F^{\theta}_l\Big(x_2,d_l\Big)\right]^T}{p^{\theta}_{U_l|T_l,{D}_{l-1}}\Big(r_l|\tau_l,d_{l-1}\Big)}dr_l\right)dx_1dx_2\right]dr_{l-1}\cdots dr_1,
\end{align}
\end{footnotesize}
and 
\begin{align}
\label{q3}
p^{\theta}_{\mathcal{U}_{l-1}|\mathcal{T}_{l-1}}\Big(r_{1:l-1}|\tau_{1:l-1}\Big) =\prod_{i=1}^{l-1}\int_{\mathbb{R}^3}f_{x_o}^{\theta}\left(r_i\right)p_{pr_i}^{\theta}\Big(x_o|\tau_i,d_{i-1}\Big)dx_o,
\end{align}
with $I^{\tau_1}_{U_1|T_1}$ given by
\begin{small}
\begin{align}
\label{initialcondition_2}
I^{\tau_1}_{U_1|T_1}(\theta)&=\int_{\mathcal{C}}\int_{\mathbb{R}^3}\int_{\mathbb{R}^3}\frac{1}{p^{\theta}_{U_1|T_1}\Big(r|\tau_1\Big)}
\begin{bmatrix} \left(df^{\theta}_{x_1}\left(r\right)\right)^T & \left(dp_{pr_1}^{\theta}\Big(x_1|\tau_1\Big)\right)^T\end{bmatrix}
\begin{bmatrix}p_{pr_1}^{\theta}\Big(x_1|\tau_1\Big) \\ f^{\theta}_{x_1}\left(r\right)\end{bmatrix}\nonumber\\
&\ \ \ \ \ \ \ \ \ \ \ \ \ \ \ \ \ \ \ \times\begin{bmatrix}p_{pr_1}^{\theta}\Big(x_2|\tau_1\Big) \\ f^{\theta}_{x_2}\left(r\right)\end{bmatrix}^T\begin{bmatrix} df^{\theta}_{x_2}\left(r\right) \\ dp_{pr_1}^{\theta}\Big(x_2|\tau_1\Big)\end{bmatrix}
dx_1dx_2dr.
\end{align}
\end{small}
\end{corollary}

\begin{remark}
Note that if the image function $q$ is independent of the parameter vector $\theta$, then, 
\begin{align*}
F^{\theta}_l\Big(x,d_l\Big)=f_x\left(r_l\right)\left(dp_{pr_l}^{\theta}\Big(x|\tau_l,d_{l-1}\Big)\right)^T,\quad x\in\mathbb{R}^3,
\end{align*}
and the expression for $I^{\tau_1,\cdots,\tau_l}_{U_l|T_l,\mathcal{D}_{l-1}}$ can be simplified as 
\begin{scriptsize}
\begin{align}
&I^{\tau_1,\cdots,\tau_l}_{U_l|T_l,\mathcal{D}_{l-1}}(\theta)\nonumber\\
&\ \ \ \ \ =\int_{\mathcal{C}}\cdots\int_{\mathcal{C}}p^{\theta}_{\mathcal{U}_{l-1}|\mathcal{T}_{l-1}}\Big(r_{1:l-1}|\tau_{1:l-1}\Big)\Bigg[\int_{\mathcal{C}} \frac{1}{p^{\theta}_{U_l|T_l,\mathcal{D}_{l-1}}\Big(r_l|\tau_l,d_{l-1}\Big)}\Bigg.\nonumber\\
&\Bigg.\ \ \ \ \ \ \ \ \ \ \ \ \times\Bigg(\frac{\partial }{\partial \theta}\int_{\mathbb{R}^3}f_{x_o}\left(r_l\right)p_{pr_l}^{\theta}\Big(x_o|\tau_l,d_{l-1}\Big)dx_o\Bigg)^T\Bigg(\frac{\partial }{\partial \theta}\int_{\mathbb{R}^3}f_{x_o}\left(r_l\right)p_{pr_l}^{\theta}\Big(x_o|\tau_l,d_{l-1}\Big)dx_o\Bigg)dr_l\Bigg]\nonumber\\
&\ \ \ \ \ \ \ \ \ \ \ \ \times dr_{l-1}\cdots dr_1\nonumber\\
&\ \ \ \ \ =\int_{\mathcal{C}}\cdots\int_{\mathcal{C}}p^{\theta}_{\mathcal{U}_{l-1}|\mathcal{T}_{l-1}}\Big(r_{1:l-1}|\tau_{1:l-1}\Big)\Bigg\{\int_{\mathbb{R}^3}\int_{\mathbb{R}^3}\Bigg[\int_{\mathcal{C}}\frac{1}{p^{\theta}_{U_l|T_l,\mathcal{D}_{l-1}}\Big(r_l|\tau_l,d_{l-1}\Big)}\Bigg.\Bigg.\nonumber\\
&\Bigg.\Bigg.\ \ \ \ \ \ \ \ \ \ \ \ \times f_{x_1}\left(r_l\right)f_{x_2}\left(r_l\right)\left(\frac{\partial p_{pr_l}^{\theta}\Big(x_1|\tau_l,d_{l-1}\Big)}{\partial \theta}\right)^T\left(\frac{\partial p_{pr_l}^{\theta}\Big(x_2|\tau_l,d_{l-1}\Big)}{\partial \theta}\right)dr_l\Bigg]dx_1dx_2\Bigg\}dr_{l-1}\cdots dr_1.
\end{align}
\end{scriptsize}
\end{remark}

\begin{proof}
See Section \ref{proof:corollary2} in the supplementary material.
\end{proof}

As mentioned in Section \ref{fundamentalsection}, for special cases of an object with a deterministic trajectory and a static object, the probability density function of the image detection process $\mathcal{G}_{\tau_1,\cdots,\tau_K}$ at fixed time points $t_0\leq\tau_1<\cdots<\tau_K$ is simplified as given by Eqs. (\ref{specific1}) and (\ref{specific2}), respectively. We next in Corollary \ref{corr3_1} to Theorem \ref{main2} calculate the Fisher information matrix for these special cases, and show that the obtained results are consistent with the results presented in \cite{2004,lin,milad1,milad2}. 

\begin{corollary}
\label{corr3_1}
For $t_0\leq\tau_1<\cdots<\tau_K$, let $\mathcal{G}_{\tau_1,\cdots,\tau_K}\Big(\left(\mathcal{U}_K,\mathcal{T}_K\right),\mathcal{C},\Theta\Big)$ be an image detection process at fixed time points $\tau_1,\cdots,\tau_K$. Assume that $p_{U_l|T_l,\mathcal{D}_{l-1}}\Big(r_l|\tau_l,\\
d_{l-1}\Big)=p_{U_l|T_l}\Big(r_l|\tau_l\Big), d_l\in\mathcal{C}^l\times\mathbb{R}^l_{[\infty]}, l=1,2,\cdots$.

1. Then, the Fisher information matrix $I_{\tau_1,\cdots,\tau_K}$ of $\mathcal{G}_{\tau_1,\cdots,\tau_K}\Big(\left(\mathcal{U}_K,\mathcal{T}_K\right),\mathcal{C},\Theta\Big)$ is given by
\begin{align*}
I_{\tau_1,\cdots,\tau_K}(\theta)=
\begin{cases}
\sum_{l=1}^KI^{\tau_l}_{U_l|T_l}(\theta), & t_0\leq\tau_1<\cdots<\tau_K,\\
0, & \mbox{otherwise},
\end{cases}
\end{align*}
where for $l=1,\cdots,K$,
\begin{small}
\begin{align*}
I^{\tau_l}_{U_l|T_l}(\theta)=\int_{\mathbb{R}^2}\frac{1}{p^{\theta}_{U_l|T_l}\Big(r|\tau_l\Big)}\left(\frac{\partial p^{\theta}_{U_l|T_l}\Big(r|\tau_l\Big)}{\partial\theta}\right)^T\left(\frac{\partial p^{\theta}_{U_l|T_l}\Big(r|\tau_l\Big)}{\partial\theta}\right)dr.
\end{align*}
\end{small}

2.1. For an object with deterministic trajectory $X_{\tau}(\theta):=\left(x_{\tau}(\theta),y_{\tau}(\theta)\right)\in\mathbb{R}^2, \tau\geq t_0$, assume that there exists an image function $q{:}\ \mathbb{R}^2\mapsto\mathbb{R}$, which describes the image of an object on the detector plane at unit lateral magnification and it is assumed to be independent of the parameter vector $\theta=(\theta_1,\cdots,\theta_n)\in\Theta$, such that 
\begin{align*}
p^{\theta}_{U_l|T_l}\Big(r|\tau\Big):=\frac{1}{M^2}q\bigg(\frac{x}{M}-x_{\tau}(\theta),\frac{y}{M}-y_{\tau}(\theta)\bigg),
\end{align*} 
where $r=(x,y)\in\mathbb{R}^2, t_0\leq\tau\leq t$, and $M>1$ is a magnification factor. Let $D_1q$ and $D_2q$ be the partial derivatives of $q$ with respect to the $x$- and $y$-coordinates, respectively. Also, let $D_jx_{\tau}$ and $D_jy_{\tau}, j=1,\cdots,n$, 
denote the partial derivatives of $x_{\tau}$ and $y_{\tau}$ with respect to the $j^{th}$ parameter coordinate, respectively. Then, for $t_0\leq\tau_1<\cdots<\tau_K$,
\begin{align*}
I_{\tau_1,\cdots,\tau_K}(\theta)=\sum_{l=1}^KI_{\tau_l}(\theta),
\end{align*}
where
\begin{small}
\begin{align*}
I_{\tau_l}(\theta)=V_{\theta}^T(\tau_l)\left(\int_{\mathbb{R}^2}\frac{1}{q(u,v)}\begin{bmatrix}(D_1q)(u,v)  \\ (D_2q)(u,v)\end{bmatrix}\begin{bmatrix}(D_1q)(u,v)  \\ (D_2q)(u,v)\end{bmatrix}^Tdudv\right)V_{\theta}(\tau_l),
\end{align*}
\end{small}
and
\begin{align*}
V_{\theta}(\tau_l):=\begin{bmatrix} 
(D_1x_{\tau_l})(\theta) & \cdots & (D_nx_{\tau_l})(\theta)
\\ (D_1y_{\tau_l})(\theta) & \cdots & (D_ny_{\tau_l})(\theta)
\end{bmatrix}\in\mathbb{R}^{2\times n}.
\end{align*}

2.2. For a static object with position $X_0(\theta)=\left(x_0(\theta),y_0(\theta)\right)\in\mathbb{R}^2$, we have, for $t_0\leq\tau_1<\cdots<\tau_K$,
\begin{align*}
I_{\tau_1,\cdots,\tau_K}(\theta)=I(\theta)=K\tilde{I}(\theta),
\end{align*}
where
\begin{small}
\begin{align*}
\tilde{I}(\theta)=V_{\theta}^T\left(\int_{\mathbb{R}^2}\frac{1}{q(u,v)}\begin{bmatrix}(D_1q)(u,v)  \\ (D_2q)(u,v)\end{bmatrix}\begin{bmatrix}(D_1q)(u,v)  \\ (D_2q)(u,v)\end{bmatrix}^Tdudv\right)V_{\theta},
\end{align*}
\end{small}
and for $\theta=\left(\theta_1,\cdots,\theta_n\right)\in\theta$,
\begin{align*}
V_{\theta}:=\begin{bmatrix} 
(D_1x_0)(\theta) & \cdots & (D_nx_0)(\theta)
\\ (D_1y_0)(\theta) & \cdots & (D_ny_0)(\theta)
\end{bmatrix}\in\mathbb{R}^{2\times n}.
\end{align*}
\end{corollary}

\begin{proof}
See Section \ref{proof:corr3_1} in the supplementary material.
\end{proof}

The material presented in Theorem \ref{main2} and Corollary \ref{corollary2} provides a mathematical framework to calculate the Fisher information matrix of image detection processes for a fixed time interval and for a fixed number of photons for a moving object with a general stochastic motion model. As mentioned before, in many biological applications, the motion of a small object in subcellular environments can be modeled by a linear stochastic differential equation. The solution of this linear stochastic differential equation can be modeled by a first order system driven by Gaussian noise. In Corollary \ref{corollary3} to Theorem \ref{main2}, we obtain recursive expressions for the Fisher information matrices for both image detection processes for a fixed time interval and fixed number of photons, in case that the dynamical system is described by a first order system with Gaussian process and measurement noise.

\begin{corollary}
\label{corollary3}
Let $\mathcal{G}^g_{[t]}\Big(\left(\tilde{X},H,\tilde{W}_g,Z_g\right),\left(\mathcal{U}_{[t]},\mathcal{T}_{[t]}\right),\tilde{\Phi},M',\mathcal{C},\Theta\Big)$ (or $\mathcal{G}^g_L\Big(\left(\tilde{X},\right.$
$\left.H,\tilde{W}_g,Z_g\right),\left(\mathcal{U}_L,\mathcal{T}_L\right),\tilde{\Phi},M',\mathcal{C},\Theta\Big)$) be an image detection process with expanded state space $\tilde{X}$ and Gaussian process and measurement noise models for a time interval $[t_0,t]$ (or for a fixed number $L$ of photons). Let $C:=M'H$. Assume that the photon detection rate $\Lambda$, $C$ and $Z_g$ are independent of $\theta$. Let
\begin{small}
\begin{align}
\label{s_theta}
&S_{\theta,l}^{(ji)}-A_{\theta,l}^{(j)}S_{\theta,l-1}^{(ji)}\left(A_{\theta,l}^{(i)}\right)^T=B_{\theta,l}^{(j)}R_{\theta,l-1}\left(B_{\theta,l}^{(j)}\right)^T,\quad l=2,3,\cdots,\nonumber\\
&S_{\theta,1}^{(ji)}=\begin{bmatrix} \tilde{\phi}_{\theta}(\tau_0,\tau_1)\bar{x}_{\theta,0}+\tilde{a}_{\theta}(\tau_0,\tau_1) \\ \frac{\partial\left(\tilde{\phi}_{\theta}(\tau_0,\tau_1)\bar{x}_{\theta,0}+\tilde{a}_{\theta}(\tau_0,\tau_1)\right)}{\partial\theta_j}\end{bmatrix}\begin{bmatrix} \left(\tilde{\phi}_{\theta}(\tau_0,\tau_1)\bar{x}_{\theta,0}+\tilde{a}_{\theta}(\tau_0,\tau_1)\right) \\ \left(\frac{\partial\left(\tilde{\phi}_{\theta}(\tau_0,\tau_1)\bar{x}_{\theta,0}+\tilde{a}_{\theta}(\tau_0,\tau_1)\right)}{\partial\theta_i}\right)\end{bmatrix}^T,
\end{align}
\end{small}
where
\begin{footnotesize}
\begin{align*}
A^{(i)}_{\theta,l}:=\begin{bmatrix} \tilde{\phi}_{\theta}(\tau_{l-1},\tau_l) & 0_{k\times k} \\ \frac{\partial \tilde{\phi}_{\theta}(\tau_{l-1},\tau_l)}{\partial\theta_i} & \tilde{\phi}_{\theta}(\tau_{l-1},\tau_l)\left(I_{k\times k}-K_{\theta,l-1}C\right)\end{bmatrix},\quad B_{\theta,l}^{(i)}:=\begin{bmatrix} \tilde{\phi}_{\theta}(\tau_{l-1},\tau_l)K_{\theta,l-1} \\ \frac{\partial \left(\tilde{\phi}_{\theta}(\tau_{l-1},\tau_l)K_{\theta,l-1}\right)}{\partial\theta_i} \end{bmatrix},
\end{align*}
\end{footnotesize}
and $R_{\theta,l}:=CP_{\theta,l}^{l-1}C^T+\Sigma_g$, $K_{\theta,l}:=P_{\theta,l}^{l-1}C^T\left(CP_{\theta,l}^{l-1}C^T+\Sigma_g\right)^{-1}, l=1,2,\cdots$, where $P_{\theta,l}^{l-1}$ is obtained through Eqs. (\ref{kalman1}) and (\ref{kalman2}).

Then, the Fisher information matrix $I_{\tau_1,\cdots,\tau_K}$ in Eq. (\ref{fixed1_1}) (or Eq. (\ref{fishi2_2})) of Theorem \ref{main2} can be calculated as
\begin{align}
\label{corol3_2}
I_{\tau_1,\cdots,\tau_K}(\theta)=
\begin{cases}
\sum_{l=1}^KI^{\tau_1,\cdots,\tau_l}_{U_l|T_l,\mathcal{D}_{l-1}}(\theta), & t_0\leq\tau_1<\cdots<\tau_K\leq t,\\
0, & \mbox{otherwise},
\end{cases}
\end{align}
where, for $\theta=(\theta_1,\cdots,\theta_n)\in\Theta$ and $l=1,\cdots,K$, the $i,j^{th},i,j=1,\cdots,n$, entry $\left[I^{\tau_1,\cdots,\tau_l}_{U_l|T_l,\mathcal{D}_{l-1}}\right]_{i,j}$ of $I^{\tau_1,\cdots,\tau_l}_{U_l|T_l,\mathcal{D}_{l-1}}$ can be calculated as 
\begin{small}
\begin{align}
\label{corol3_3}
\left[I^{\tau_1,\cdots,\tau_l}_{U_l|T_l,\mathcal{D}_{l-1}}(\theta)\right]_{i,j}=\frac{1}{2}\mbox{trace}\left[R_{\theta,l}^{-1}\frac{\partial R_{\theta,l}}{\partial\theta_i}R_{\theta,l}^{-1}\frac{\partial R_{\theta,l}}{\partial\theta_j}\right]+\mbox{trace}\left\{R_{\theta,l}^{-1}\tilde{C}S_{\theta,l}^{(ji)}\tilde{C}^T\right\},
\end{align}
\end{small}
with $\tilde{C}:=\begin{bmatrix} 0_{2\times k} & C \end{bmatrix}$.
\end{corollary}

\begin{proof}
See Section \ref{proof:corollary3} in the supplementary material.
\end{proof}

In Section \ref{sup:ex2}, we provide an example to illustrate our results for calculating the Fisher information matrix for the specific case  of a linear trajectory described in the example provided in Section \ref{sup:ex1} of the supplementary material.  

\subsection{CRLB and standard deviation of estimates for different photon counts}

\begin{figure}[htbp]
\centering\includegraphics[width=1\textwidth]{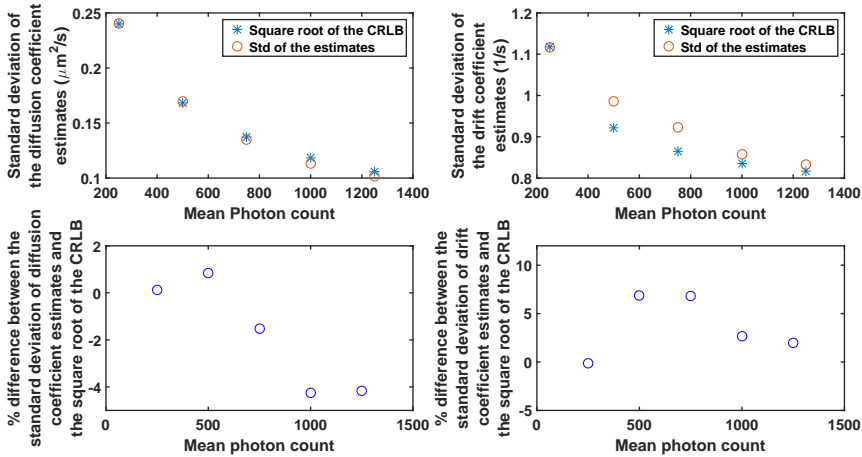}
\DeclareGraphicsExtensions{.eps} \caption{Analysis of the standard deviation of diffusion coefficient and first order drift coefficient estimates produced by the maximum likelihood estimation method for the Gaussian measurement noise case. Shown in the first row are the standard deviations of the diffusion coefficient and first order drift coefficient estimates versus the square roots of their corresponding CRLBs for simulated data sets. The simulated data sets are the detected photons emitted from a molecule, referred to as the images of a molecule, with a stochastic trajectory which differ by the mean photon count assumed for each trajectory. For each mean photon count, the data set consists of 100 repeat images. For a given data set, the time points of the detected photons are drawn from a Poisson process and are the same for the all trajectories. All trajectories are simulated in the object space using Eq. (\ref{linearspecific}) with the first order drift coefficient $F=-10I_{2\times 2}$/s and the diffusion coefficient $D=1$ $\mu\mbox{m}^2$/s. We assume the zero order drift is equal to 0. Also, we assume that the initial location of the molecule is Gaussian distributed with mean $x_0=(5,5)^T$ $\mu$m and covariance $P_0=10I_{2\times 2}$ n$\mbox{m}^2$. Detected locations of the photons emitted from the molecule in the image space are simulated using Eq. (\ref{observeEq_gaussian}) with the parameters given in Section \ref{parameters}. Shown in the second row are the percentage differences between the standard deviation of the diffusion coefficient and first order drift coefficient estimates and the square roots of their corresponding CRLBs.}
\label{thirdimage}
\end{figure}

We next evaluate the performance of our proposed maximum likelihood estimation method in terms of the standard deviation of the estimates. For this purpose, we simulated data sets of the detected photons emitted from a molecule, referred to as the images of a molecule, with a stochastic trajectory which differ by the mean photon count, i.e., the mean number of detected photons during the exposure time interval, assumed for each trajectory. This mean photon count ranges from 250 to 1250. For each mean photon count, the data set consists of 100 repeat images simulated using the Gaussian profile (Eq. (\ref{observeEq_gaussian})) with the parameters given in Section \ref{parameters}. For these data sets, we calculated the maximum likelihood estimates of the diffusion and first order drift coefficients, separately. Also, for the given data set and time points, we obtained the square roots of the CRLBs for the diffusion and first order drift coefficient by calculating the square roots of the inverse of their corresponding Fisher information matrices at the fixed time points. It can be seen in the first row of Fig. \ref{thirdimage} that as the mean photon count increases and thereby the amount of data that is available for the estimation increases the CRLB for the estimates improves, consistent with the expectation that an increasing amount of data leads to improved estimation results. The standard deviations of the estimates show the analogous behavior while exhibiting the expected fluctuations due to the stochastic nature of the sample standard deviations. Also, the percentage differences between the standard deviations and the square roots of the CRLBs are shown in the second row of Fig. \ref{thirdimage}. The percentage difference is the difference between the standard deviation of the estimates and the square root of the corresponding CRLB, expressed as a percentage of the square root of the corresponding CRLB. As can be seen, these percentage differences are at most around 10\%.

Note that in theory, the square root of the CRLB provides a lower bound on the standard deviation of an unbiased estimator. However, in simulation results,  we deal with individual stochastic trials rather than the probabilistic expressions that are used in the statements of the CRLB. This means that even if the (probabilistic) standard deviation of an estimator attains the CRLB, the sample standard deviation obtained in a stochastic simulation will deviate from the probabilistic expression and could be expected to be both above and below the CRLB.

\subsection{Fisher information matrix for non-Gaussian measurement noise}

So far, for computational purposes and taking advantage of the Kalman filter formulation, we have focused on computing the Fisher information matrix and CRLB only for Gaussian measurements. Although the Gaussian assumption is very useful in some applications, there are many cases for which this assumption can be problematic in practice due to the fact that the Gaussian model is often not a suitable approximation for an analytical image profile. As mentioned earlier, from optical diffraction theory, a typical point spread function for an in-focus molecule is given by the Airy profile. Also, for the out-of-focus scenario, the image function is given by a classical model of Born and Wolf \cite{born}.

\begin{figure}[htbp]
\centering\includegraphics[width=1\textwidth]{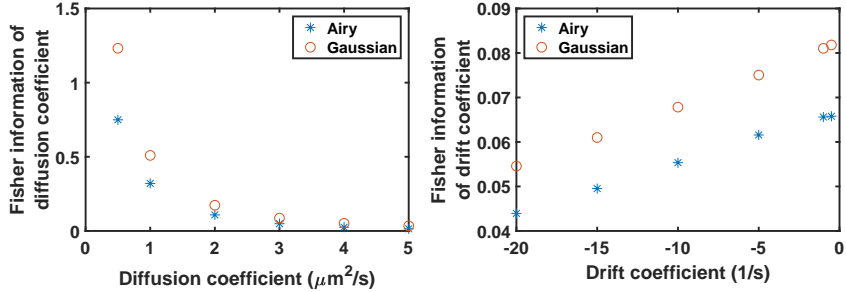}
\DeclareGraphicsExtensions{.eps} \caption{Fisher information matrix for Airy measurement noise versus Gaussian measurement noise. Fisher information matrix of diffusion and first order drift coefficients for the Airy measurement noise with parameter $\alpha=2\pi n_a/\lambda$ given in Section \ref{parameters} and by a 2D Gaussian profile with standard deviation $\sigma=1.323/\alpha$, in case that we have one photon with arrival time of $\tau_1=20$ ms.}
\label{airy_gauissan}
\end{figure}

Here, we computed the Fisher information matrix of both the first order drift and diffusion coefficients for the Airy measurements case and compared the results with the Fisher information matrix obtained for the case that the Airy profile is approximated by a 2D Gaussian profile. The typical approximation of the Airy profile with $\alpha:=2\pi n_a/\lambda$ by a 2D Gaussian profile with standard deviation $\sigma$ yields a value of $\sigma=1.323/\alpha$ \cite{2004}. We only focused on the one photon case, since computing the integrals of the Fisher information expression for the Airy profile case numerically requires a large number of samples and it is computationally expensive (see Section \ref{fisherproc} in the supplementary material for the detailed computational procedure). As shown in Fig. \ref{airy_gauissan}, the difference between the Fisher information matrices of these two different profiles can be significant. As can be seen, this difference goes to 0 for the diffusion coefficient and remains constant for the estimation of the drift coefficient. The other main difference is that the CRLB decreases in the diffusion estimation case, whereas it increases for the drift estimation as the size of the drift increases.

\subsection{CRLB and Fisher information matrix for different sets of time points}

\begin{figure}[htbp]
\centering\includegraphics[width=1\textwidth]{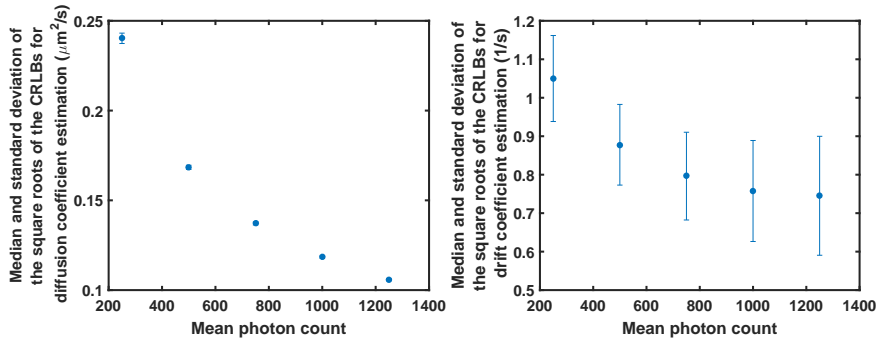}
\DeclareGraphicsExtensions{.eps} \caption{Analysis of the square root of the CRLB of the diffusion coefficient and first order drift coefficient estimates for different sets of Poisson distributed time points. Medians and standard deviations of the square roots of the CRLBs of the diffusion coefficient and the first order drift coefficient estimates are shown by the circles and error bars, respectively, for the simulated trajectories with the same parameters as in Fig. \ref{thirdimage}, and different time points drawn from a Poisson process with the same mean value, which ranges from 250 to 1250.}
\label{fourthimage}
\end{figure}

To examine further the CRLB for parameter estimation for a moving single molecule with a stochastic trajectory, we calculated the square root of the CRLB for the simulated trajectories with the same parameters as in Fig. \ref{thirdimage}, and different time points drawn from a Poisson process with a mean value which ranges from 250 to 1250. In Fig. \ref{fourthimage}, we have plotted, for a given photon count, the median and standard deviation of the different simulations of the square root of the CRLB for both the diffusion and drift estimates. As in prior analyses the results vary significantly for the two scenarios. While in both cases the median decreases with increasing photon count (and thereby increasing acquisition time), the standard deviations of the square roots of the CRLB expressions behave very differently. In case of the diffusion parameter the standard deviations of the CRLB expressions are almost insignificant, indicating that the specifics of the photon detection times do not have a major impact on the standard deviation with which the diffusion coefficient can be estimated. The situation for the estimation of the drift parameter is, however, very different. Here the corresponding standard deviations are relatively high and in fact increase with the number of photons that are acquired. This shows that the standard deviation with which the drift coefficient can be estimated, in contrast to the diffusion coefficient, is highly dependent on the specific time points at which the emitted photons are detected.

\begin{figure}[htbp]
\centering\includegraphics[width=1\textwidth]{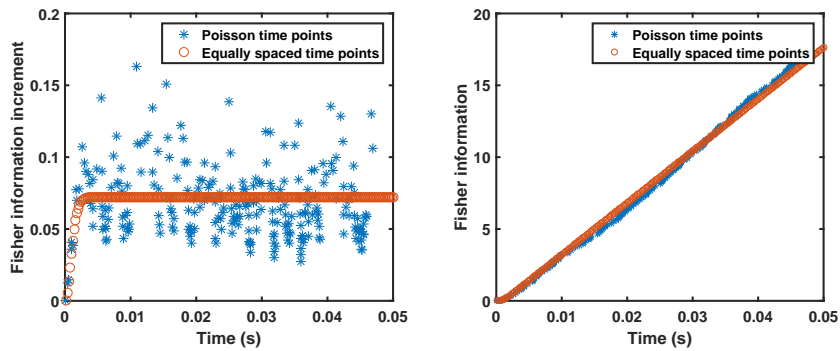}
\DeclareGraphicsExtensions{.eps} \caption{Fisher information analysis of single molecule trajectories simulated using Poisson distributed and equally spaced time points. Shown in the left are the Fisher information matrix increments on the diffusion coefficient estimation for data sets of two trajectories, first containing a trajectory of a molecule simulated using Eqs. (\ref{linearspecific}), where the time points are drawn from a Poisson process with mean 250 in the time interval $\left[0,50\right]$ ms, and second containing 250 equally spaced time points in the time interval $\left[0,50\right]$ ms, with the parameters given in Fig. (\ref{thirdimage}). Shown in the right is the Fisher information matrix (sum of the increments) for both trajectories.}
\label{fimAnalysis}
\end{figure}

We also show the Fisher information matrices (and Fisher information matrix increments, i.e., the amount of information obtained by detecting one additional photon) for Poisson distributed time points and for equally distributed time points in Fig. \ref{fimAnalysis}. For this purpose, we simulated two data sets of single molecule trajectories with Gaussian measurements, the first containing a trajectory of a molecule simulated using Eqs. (\ref{linearspecific}), where the time points are drawn from a Poisson process with mean 250 in the time interval $\left[0,50\right]$ ms, and the second containing 250 equally spaced time points in the time interval $\left[0,50\right]$ ms. We then calculated the Fisher information matrix increments and Fisher information matrix (sum of the increments) on the diffusion coefficient estimation for both data sets. As can be seen, the Fisher information matrix increments, after initial iterations, are constant for the case of equally spaced time points. However, for different realizations of Poisson time points, the Fisher information matrix increments are different from each other. When the time difference between two successive time points decreases (increases), the Fisher information matrix increment of the diffusion coefficient decreases (increases), and conversely, the corresponding CRLB increases (decreases).  

\bibliographystyle{siamplain}
\bibliography{references}
\end{document}